\documentclass[%
reprint,
superscriptaddress,
groupedaddress,
showpacs,preprintnumbers,
aps,
prb,
]{revtex4-2}

\usepackage{amsmath}
\usepackage[]{graphicx}
\usepackage{makeidx}
\usepackage{verbatim}
\usepackage[colorlinks,urlcolor=blue,linkcolor=blue,citecolor=blue]{hyperref}
\usepackage{dcolumn}
\usepackage{bm}
\usepackage{multirow}
\usepackage{float}
\usepackage{booktabs}
\usepackage{tabularx}
\usepackage{makecell}
\usepackage{siunitx}
\usepackage{array,mathtools,amssymb,booktabs,longtable}
\usepackage[caption=false,subrefformat=parens,labelformat=parens,listofformat=subparens]{subfig}
\usepackage{adjustbox}
\usepackage{amsfonts}
\usepackage{ragged2e}
\usepackage[dvipsnames]{xcolor}
\usepackage{verbatim}
\usepackage[normalem]{ulem}

\usepackage{braket}

\newcolumntype{C}{>{$}c<{$}}
\AtBeginDocument
{
	\heavyrulewidth=.08em
	\lightrulewidth=.05em
	\cmidrulewidth=.03em
	\belowrulesep=.65ex
	\belowbottomsep=0pt
	\aboverulesep=.4ex
	\abovetopsep=0pt
	\cmidrulesep=\doublerulesep
	\cmidrulekern=.5em
	\defaultaddspace=.5em
}

\graphicspath{{figures/}}

\newcommand{\mbb}[1]{\mathbb{#1}}

\begin{document}
	\title{Topological electronic structure of twin boundaries \\ and twinning superlattices in the SnTe material class}
	\author{Saeed Samadi}
	\author{Rafa\l{} Rechci\'{n}ski}
	\author{Ryszard Buczko}
	\email[]{buczko@ifpan.edu.pl}
	\affiliation{Institute of Physics, Polish Academy of Sciences, Aleja Lotnik\'{o}w 32/46, 02668 Warsaw, Poland}
	
	\date{\today}
	\begin{abstract}
		The topological electronic structure of a single twin boundary and coherent twinning superlattices (TSLs) based on the SnTe class of material is calculated and discussed within a supercell implementation. The superlattices consist of two twin planes (TPs) in the supercell arranged in such a way that each of the boundaries forms a mirror plane for the entire structure. Two types of TP boundary, cationic and anionic, can exist, and so three types of supercells can be constructed. We study the topological phases of each twinning configuration using the tight-binding approximation and calculating the topological invariants. We show that they differ by topological properties and find that all-cationic TSLs are topologically distinct from the anionic case due to the opposite sign of the Berry curvature around the $ \rm \Gamma $ point of the TSLs Brillouin zone. Our findings are consistent with a complementary analysis of (111)-oriented slabs with a single twin boundary in the presence of the Zeeman field. They are also consistent with the calculated number of spin-polarized Dirac-like edge states of both superlattices and slabs. We conclude that each type of TP forms the 2D mirror-plane-protected topological crystalline insulator. 
		\end{abstract}
	\maketitle
	
	\section{\label{sec:intro} Introduction}

	Topological crystalline insulators (TCIs) are materials in which crystal symmetry protects the nontrivial topology of the electronic band structure \cite{Fu2011TCI}. In IV-VI TCIs the the topology is protected by $ \{110\} $ mirror symmetries and negative bulk band gap leads to Dirac-like metallic surface states \cite{hsieh2012topological, Ando_TCI, Wang2016, safaei2013, Liu2013, Buczko2020}. Breaking the crystal symmetry generates a Dirac mass and gives rise to gapped phases \cite{hsieh2012topological, Okada2013science, zeljkovic2015dirac} with potentially novel functionalities in low-power electronics and spintronics \cite{Liu_2014spin}. Such a prospect motivates the design of nanostructures based on IV-VI TCI semiconductors allowing for control of the number of carriers and electronic states by tunable gate voltage. There are many theoretical studies of single atomic layers, thin films, and heterostructures grown in (001) and (111) crystallographic directions, where the nontrivial topology is manifested by a mirror Chern number and $ \mathbb{Z}_2 $ strong and weak invariants, respectively \cite{Wrase2014, Liu2015, Niu2015, Liu_2014spin, Ozawa_2014, safaei2015quantum, Liu_2015, Yang2014hetero}. Also, some work has been involved to propose a platform for investigating hinge states \cite{schindler2018higher} and Majorana bound states \cite{Sadowski_2018,Liu_2021NWs,nguyen2022NW} in [001] oriented nanowires.

	In addition to clean surfaces of bulk material and edges of nanostructures, certain lattice defects in topologically nontrivial materials are also known to bind topological states\cite{teo2017topological}. In the case of IV-VI TCIs, anomalous helical modes bound to disclinations were predicted by theory \cite{Lau2021}. Furthermore, one-dimensional (1D) gapless modes bound to atomic step edges on the (001) surface were discovered in experiment \cite{sessi2016robust}. Their presence is attributed to emergent nontrivial topology of the 2D surface bands \cite{sessi2016robust,rechcinski2018topological,Iaia2019,Brzezicki2019}.
	
	 In this work we extend the research of the effects of lattice defects on the electronic structure topology to twin planes (TPs), that is, planar defects resulting from stacking faults during growth in a fixed crystallographic direction. In cubic crystals, most of the faults appear predominantly in the $ [111] $ direction without breaking any bonds. They are characterized by zero stress and have very low formation energies. On the atomic scale, these unique features ensure the maximum degree of symmetry and coherence compared to other types of grain boundaries. In particular, these are defects of a different type than grain boundaries formed at the border of two crystals twisted relative to each other at a small angle and investigated in the context of non-trivial systems at the interface between topological insulators \cite{Slager2016, Slager2019}.  TPs are known to be commonly observed in group IV (e.g., Si) and III-V (e.g., GaAs and InP) semiconductor nanowires. In IV-VI semiconductors, twinning can be observed as interfaces between crystallites, for example, in the mineral galena, i.e., the rocksalt-type PbS \cite{galena}. TPs have been observed also in $ \mathrm{Pb_{1-x}Sn_{x}Te} $ NWs grown along the $[011]$ direction. The wires have a pentagonal cross section with five $\left\lbrace 111 \right\rbrace $ twins extended radially from the center of the wire \cite{Janaszko_2020_poster}. Some theoretical attempts towards electronic properties of twin boundaries in rock-salt crystals have been reported in Refs.~\cite{Ikonic1995twin,Ikonic2001electronic}. 
	 
	 Interestingly, also periodically arranged TPs can be found among natural minerals. In mineralogy, they are known as polysynthetic twins and may appear in a number of crystals, notably plagioclase feldspars, such as albite \cite{Sorrel1977}.
	 Such twinning superlattices (TSLs), have drawn a great deal of attention in the past several decades and have been explored as a new class of crystal structure in cubic and zinc-blende binary compounds in semiconductors and metals \cite{Ikonic1995twin, Ikonic1993}. Therefore, some efforts have been made both theoretically and experimentally to understand and control twinning processes on the nanoscale, making it possible to fabricate nanoengineered twinning superlattices in crystalline nanowires \cite{Porter2016thermal,Ghukasyan2022phase,algra2008twinning, wood2012growth}.  
	
	 In order to study the topological properties of TPs in IV-VI TCIs we choose cubic SnTe as a model representative of a whole class of similar materials. In addition to SnTe  it includes in particular  (Pb,Sn)Se, (Pb,Sn)Te, cubic SnSe, and SnS. Depending on the pressure, composition, temperature and doping, they have rock salt, rhombohedral (tellurides) or orthorhombic (selenides) structure \cite{Strauss1967, Brillson1974, Littlewood1980, Murase1981, Plekhanov2014}. SnSe and SnS can crystallize in a cubic structure, although it is metastable under normal conditions \cite{Sun2013, Mariano1967, Wang2015, Jin2017, Sihi2022}. The energy gap in binary compounds is inverted. In crystalline solid solutions of SnSe and SnTe with lead, the gap depends on the composition and external conditions \cite{Dimmock1966, Martinez1973a, Martinez1973b, Martinez1973c,  Wojek2014, Neupane2015}.  
	 
	 We start with a theoretical study of hypothetical TSLs with rock-salt structures grown along the [111] direction. Then we turn to the study of a single twin boundary present in the middle of (111)-oriented slabs. The vicinity of a single (111) TP can be viewed as composed of two perfect crystals, which are rotated around $[111]$ by $180^{\circ}$ with respect to each other and are joined in this plane. Structurally, it can be described as the reversal of atomic stacking sequence along the [111] direction.  Although the interface between the two [111] crystal orientations is perfectly lattice-matched, the wave functions are highly symmetry-mismatched. This makes the twin stacking fault, in a sense, a junction of two essentially different materials, even though the material is of the same composition and lattice type on both sides. To quantify the effect of TPs on the electronic band structure we use tight-binding (TB) calculations. We have found that TPs significantly affect the electronic properties of SnTe, including the topology of the band structure. We show that the topological features of the studied models are strongly influenced by the types of TPs present in the structure. 
	 
	 By symmetry, the unit cell of each TSL must contain an even number, at least two, TPs. In contrast, slabs have single TPs, but also surfaces with nontrivial electronic states. The combination of the results obtained for both types of structures allows us to conclude that each individual TP can be treated as a two-dimensional topological crystalline insulator with corresponding nontrivial states along its edges.

	\begin{figure}[!htbp]
		\centering
		\includegraphics[width=0.48\textwidth]{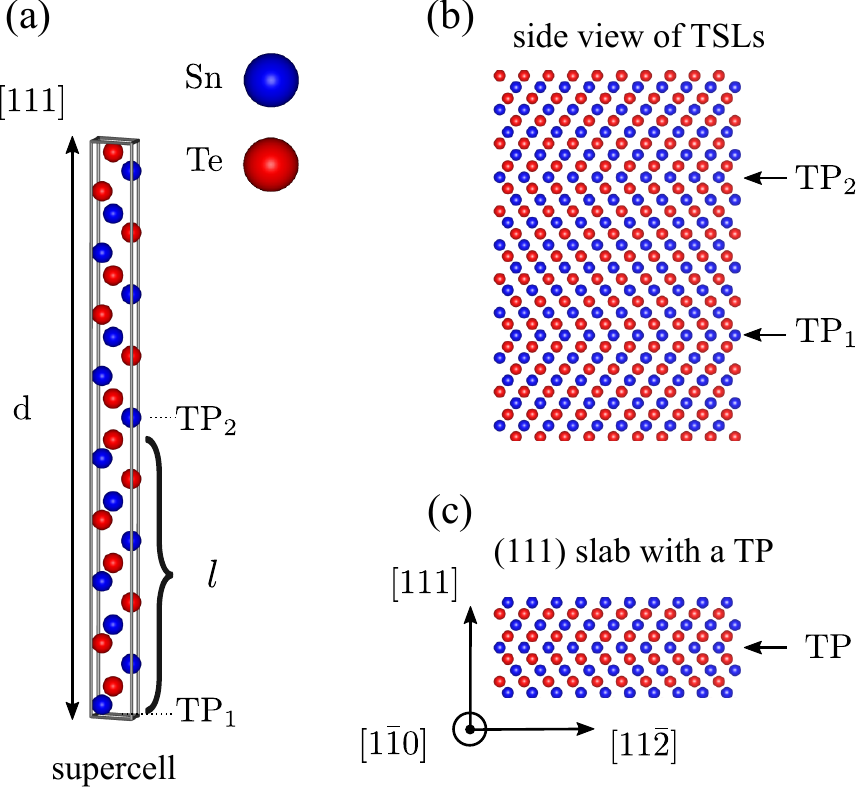}
		\subfloat{\label{fig:str_supercell}}				
		\subfloat{\label{fig:str_TSLs}}				
		\subfloat{\label{fig:str_slab}}				
		\caption{\footnotesize
			 (a) [111] primitive supercell with height $ d=28 $ ($ l=14 $) atoms and (b)  $ (1\bar 1 0) $ side view of a TSL 3D bulk.  (c) Sketch of (111)-oriented slab structure with thickness of 9 atomic layers with a single TP in the middle. }\label{fig:str}
	\end{figure}

	\section{\label{sec:methods} Computational methodology }

	The calculations were performed in the framework of tight-binding (TB) approximation. We used a simplified model in which only $p$ orbitals with $\sigma$ bonds between nearest and next-nearest neighbors were considered. The Hamiltonian for such a model was proposed by Hsieh et al.~\cite{hsieh2012topological}, and appropriate parameters were published in~\cite{Fulga2016}. In our calculations we used spin-orbit parameter $\lambda=0.3$ instead of $\lambda=0.7$ in order to reduce band gap to $0.33$ eV, which is close to SnTe gap in the ambient conditions. 
	Such model captures all essential qualitative aspects of the TCI bulk spectrum while allowing efficient large-scale calculations.
	
	 We first study three-dimensional TSLs in which the supercell includes two $ (111) $ planar faults (i.e., TPs) and the structure is grown along the [111] direction.  However, the atomic stacking sequence is changed due to the presence of a TP in such a way that creates a local (111) reflection symmetry with respect to each TP. Because a defect can appear on either an anionic or a cationic layer depending on which type of atoms occupy the twinning plane, we are dealing with three different types of supercells: having two cationic TPs (cat-cat), two anionic TPs (an-an), and a cationic and an anionic TP (cat-an). These correspond to structures with all cationic, all anionic, and alternating cationic and anionic TPs, respectively. The supercell and TSLs as schematically depicted in Fig.~\subref*{fig:str_supercell}-\subref{fig:str_TSLs} are built symmetrically with respect to each TP, and the atomic height of the supercell ($ d $) along [111] can be adjusted by setting a given atomic distance ($ l = d/2 $) between two twin boundaries.  The number of atoms in the supercell $d$ is always even. For supercells in which $ l $ is even, the atomic type of TPs is the same. Conversely, for odd $ l $ the TPs are of different types. Thus, cat-cat and an-an TSLs are identified by the space group $P6_3/mmc$ ($ D_{6h}^4 $, No.194), while cat-an TSLs have a different space group denoted $ P\bar6 m2 $ ($ D_{3h}^1 $, No.187) \cite{Bilbao}. We point out that these space groups formally belong to the hexagonal lattice system; however, following Refs.~\onlinecite{safaei2015quantum} and \onlinecite{Liu_2015}, when referring to crystal planes and directions in this work, we use the Miller indices for a cubic lattice, according to the lattice vectors of an ordinary defect-free SnTe far away from the TPs.

	Of the many symmetries of TSL systems, only a few are significant in this study.  In addition to (111) reflection symmetry mentioned above, we emphasize that there exists an inversion symmetry in the TSLs with TPs of the same kind, whereas it is absent in the cat-an TSL structures. Inversion, together with time-reversal symmetry, leads to at least a twofold degeneracy of the energy bands for cat-cat/an-an TSLs. Also, it is worth highlighting that the $ C_3 $ rotation symmetry as well as the $ \{110\} $ mirror symmetry are present for all cases.

\begin{figure}[!htbp]
		\centering
		\includegraphics[width=0.48\textwidth]{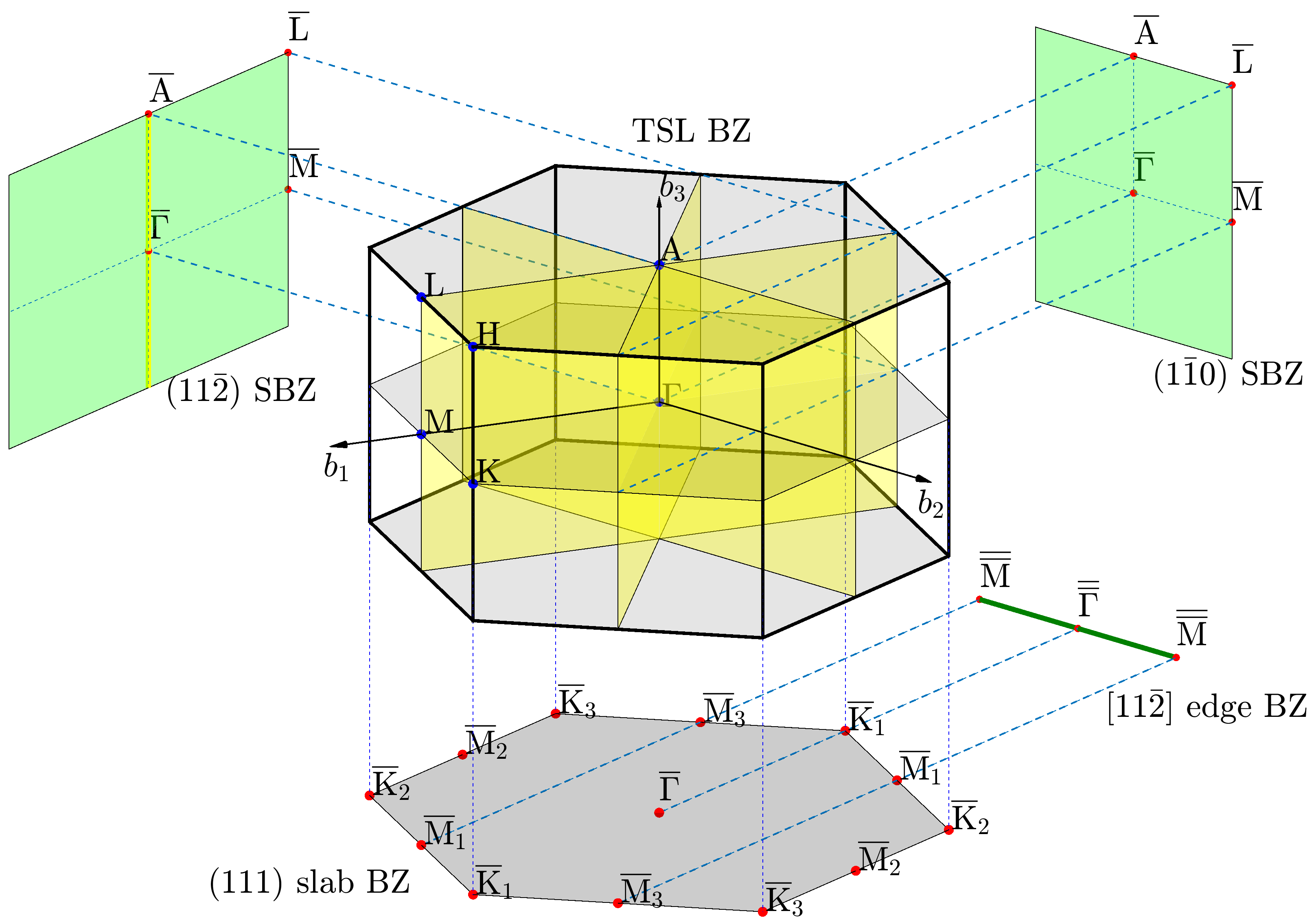}
		\caption{\footnotesize
			 Three dimensional simple hexagonal Brillouin zone of a [111]-oriented TSLs (center) with the corresponding surface Brillouin zones (SBZ) (green). The yellow-shaded rectangles denote $ \{110\} $ mirror symmetry planes, and the light grey-shaded hexagons indicate (111) mirror planes. (bottom) The 2DBZ of (111)-oriented slab (gray hexagon) is shown with 1DBZ of its $ [11\bar 2] $ edge (green line). }\label{fig:bz_hex}
	\end{figure}

	The TSL structure has a simple hexagonal Brillouin zone (BZ), as shown in Fig.~\ref{fig:bz_hex}. The height of the TSL BZ along the [111] direction is inversely proportional to the height of the supercell. The yellow-shaded planes inside the BZ denote $ \{110\} $ mirror planes. Of six such mirror planes in a bulk defect-free SnTe, only three of them are left in the TSLs. 
	
	Furthermore, in order to study lateral surface states, we consider TSLs truncated along $ (1\bar 1 0) $ and $ (1 1 \bar 2) $ faces, which are perpendicular to the (111) TPs. The surface Brillouin zones (SBZ) corresponding to these orientations are depicted in light green in Fig.~\ref{fig:bz_hex}. In the 3D hexagonal BZ there exist three different $\rm   L $ ($\rm   M $) points, where for both surface orientations two of them are projected into the same point labeled $\rm  \overline{ L} $ ($\rm  \overline{ M} $), respectively. The third high-symmetry point $\rm   L $ ($\rm  M $) is projected together with  $\rm  A $  ($ \rm \Gamma $) onto $\rm \overline{A}$ ($\overline{\rm \Gamma}$). In the case of $ (1 \bar 1 0) $ SBZ, the points $\rm  H $ and $\rm   K $ also project onto $\rm \overline{A}$ and $\rm \overline \Gamma$, respectively.  Note that the $(11\bar2)$ surface preserves one of the $ \{110\} $ mirror planes, whereas the $(1\bar10)$ surface breaks all of them. Both surface terminations preserve the $(111)$ mirror plane.

	We recall that surfaces of defect-free SnTe feature nontrivial states in the vicinity of the projections of the $\rm L^{\rm fcc}$ points of the bulk BZ onto the SBZ \cite{safaei2013} (the use of superscript $\rm fcc$ is to distinguish the point from the $\rm L$ points in the hexagonal BZ). In the case of the TSL supercell, since the height $d$ can be tuned by any even integer number, it turns out that the $\rm   L^{\rm fcc} $ points reduced to the TSL BZ are not generally unique and may fall on different points, that are always lying on the $\rm \Gamma$-$\rm A$, and $\rm M$-$\rm L$ lines. 
	
		Moreover, to calculate the topological properties of a single TP we use models of (111) slabs schematically demonstrated in Fig.~\subref*{fig:str_slab}. TP is situated in the middle of each slab and forms (111) mirror plane symmetry. The point group symmetry of the structure is $D_{3h}$, however, the 2D space group is $P3m1$. The corresponding 2D hexagonal BZ of the slab (gray) is shown with 1DBZ of its $ [11\bar 2] $ edge in Fig.~\ref{fig:bz_hex}. The $ \rm \overline{\overline{M}} $ point, in this particular choice, represents the projection of $ \rm \overline{ M}_1 $ and $ \rm \overline{ M}_3 $, while the $ \rm\overline{\overline{\Gamma}} $ point shows the projection of $\rm \overline{ M}_2 $ and $ \rm\overline{ \Gamma} $ in its 1DBZ.

	\section{\label{sec:results} Results and discussions}

	\subsection{\label{subsec:TSL}Twinning superlattices}
	
	\subsubsection{\label{sec:bulk_bs}\textbf{ Electronic band structure and topological invariants}}
	
	\begin{figure}[!htbp]
		\centering
		\includegraphics[width=0.48\textwidth]{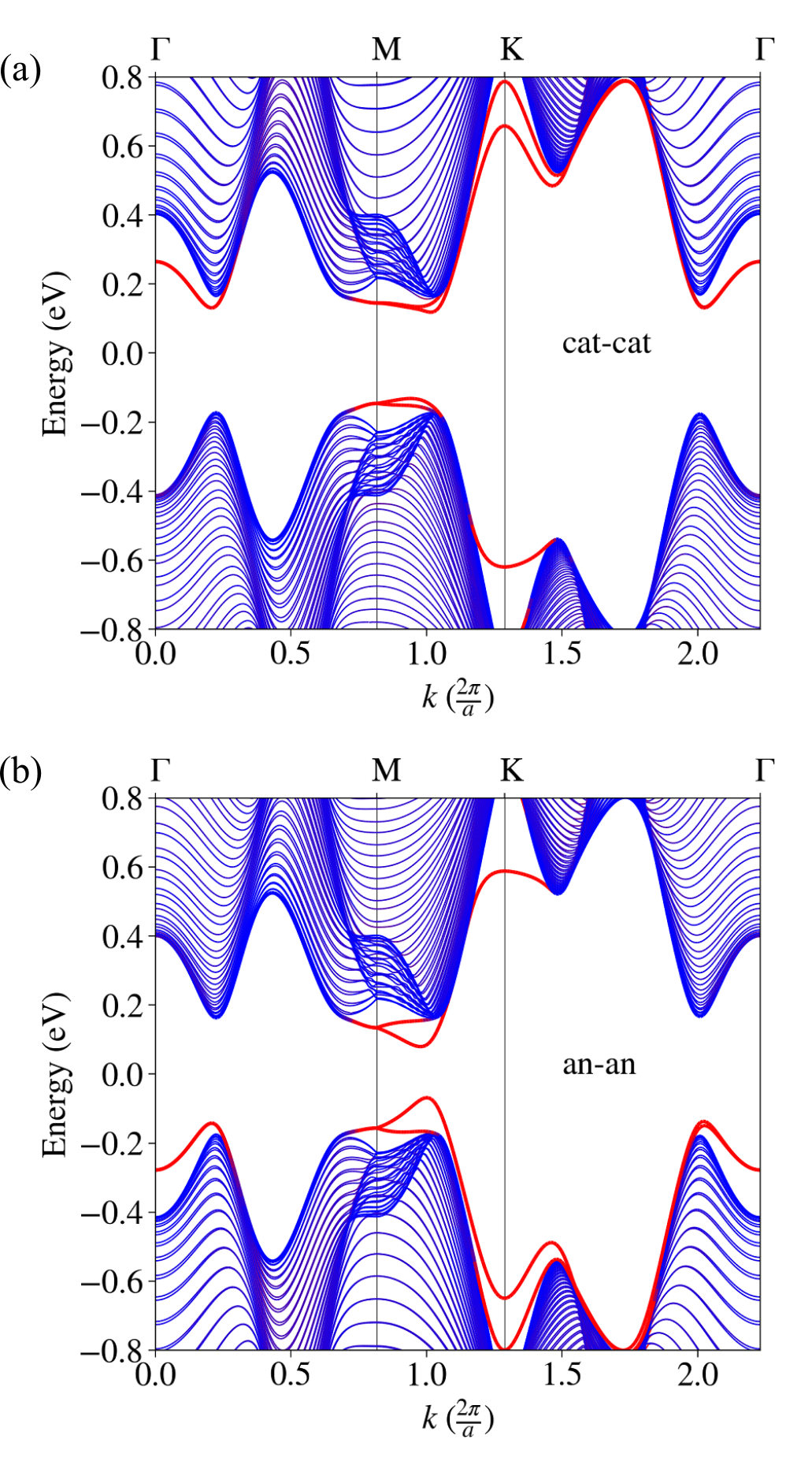}
		\caption{\footnotesize   The calculated band structures of 200 monolayer-height [111]-oriented SnTe TSLs for the $ k $  wave vectors along $ \rm \Gamma $-$\rm  M $-$\rm  K $-$ \rm \Gamma $ high symmetry lines of the BZ. The TSLs  host (a) all-cationic TPs and (b) all-anionic TPs and the red to blue color coding indicates the localization of the states at TPs and at  away from the TPs.}\label{fig:bs_t11_t22}
	\end{figure}

	In this section, we present band structures of TSLs for various TP configurations. 
	
	The electronic band structures are calculated along high-symmetry lines demonstrated in Fig.~\ref{fig:bz_hex}. In particular, in Fig.~\ref{fig:bs_t11_t22} we show the results for an-an and cat-cat TSLs along the $ \rm\Gamma$-M-K-$\rm \Gamma $ path. We inspect the influence of TPs by coloring the spectral lines according to the localization of the wave functions with the red color denoting close proximity to the TPs. The results show that the states localized on the TPs are situated in the bulk band gap region near the $\rm M $ points. Moreover, the localized states are revealed in the conduction (valence) band edge around the $ \rm \Gamma $ point for cat-cat (an-an) superlattices. We considered a supercell with $d=200$ atomic layers (i.e., the distance between two twin planes is $l=100$ monolayers), a height large enough to suppress any hybridization between the states localized at TPs, i.e., further increase of $d$ has no effect on the localized states. Since cat-cat and an-an TSLs preserve both time-reversal and inversion symmetries, all bands must have even degeneracy by the Kramers theorem. The band edges depicted with the red lines in Fig.~\ref{fig:bs_t11_t22} are twofold degenerate in the $\rm \Gamma M K$ plane.
	
	For each type of TSL we also calculate topological invariants: mirror Chern numbers $C_m$ and the $\mathbb{Z}_2$ strong and weak indices $(\nu_0;\nu_1,\nu_2,\nu_3)$. We evaluate $C_m$ with respect to mirror planes $(111)$ and $(1\bar10)$, the former on two different mirror-reflection invariant planes $\mathrm{\Gamma M K}$ and $\mathrm{A L H}$, and the latter on the plane $\mathrm{AL\Gamma M}$. Due to $C_3$ symmetry, the result for $(1\bar10)$ is the same as for $(01\bar1)$ and $(\bar101)$.

	For the calculations of $C_m$ and $\mbb Z_2$ we used the numerical schemes due to Fukui et al.~\cite{FukuiChern,FukuiZ2}, except for $\mbb Z_2$ in inversion symmetric supercells for which Fu and Kane's method was used~\cite{FuKane2007,FuKaneMele2007}. We perform the calculations for supercell heights $d\geq8$ for cat-cat/an-an TSLs and $ d\geq10 $ for cat-an TSLs. The evaluated values of $ C_m $ and $(\nu_0;\nu_1,\nu_2,\nu_3)$ are shown in Table~\ref{T1}, for three different TSL configurations. The weak indices $ \{\nu_i,i=1,2,3\} $ characterize the topology of appropriate time-reversal invariant planes in the BZ. $\nu_3$ in particular describes the ALH plane.

	Excluding the smallest possible width $d$ (i.e., cat-cat / an-an: four-layer height and cat-an: six-layer height), the topological indices stay constant and do not change with increasing width $d$. Note that the invariants converge to constants at $d=8$ or $d=10$ not universally, but only in this particular simplified TB model. These thresholds will be different for different models, but the fact that they converge at some $d$ should be universal.  Thus, the calculated mirror Chern number in the $ \rm \Gamma MK $ plane is equal $ C_m^{(\rm \Gamma M K)} =4 $ for cat-cat TPs in the supercell, while for an-an TSLs $ C_m^{(\rm \Gamma M K)} =2 $. In both stated cases the mirror Chern number for the $ \rm A LH $ reciprocal plane attains zero value. In contrast to all cationic and all anionic TSLs that exhibit \emph{even} mirror Chern numbers, for the cat-an TSL the mirror Chern number is \emph{odd} for both $\rm \Gamma M K $ ($ C_m^{(\rm \Gamma M K)} =3 $) and $ \rm ALH $ ($ C_m^{(\rm A L H)} =1 $) planes. Considering the $ \rm \Gamma M K $ plane, we conclude that three TSL configurations belong to distinct topological classes since their mirror Chern numbers are different.
	
	Additionally, the $ C_m ^{(\rm A L \Gamma M)} $ corresponds to the $ (1\bar 1 0) $ mirror plane. Consistent with the defect-free bulk crystal\cite{hsieh2012topological}, the Chern number on this plane is equal to $ C_m ^{(\rm A L \Gamma M)}=2 $  (band structures  on the $\rm A L \Gamma M $ plane are provided in Appendix~\ref{app:BS}).
	
	Due to the even parity of Chern numbers in cat-cat and an-an TSLs, the $ \mbb Z_2 $ topological indices are $ (0;0,0,0) $. On the contrary, cat-an TSLs are weak topological insulators with indices $(0;0,0,1)$.

	\begin{table}[!tbp]
		\caption{ \footnotesize
			The mirror Chern number $ C_m $  and $ \mathbb{Z}_2 $  invariants calculated for 3D TSLs grown along [111] crystallographic direction with various heights of supercells. The Chern numbers $C_m$ were evaluated at three different high-symmetry planes in the BZ, indicated in the superscripts.}
		
		\begin{center}
			\renewcommand{\arraystretch}{1.5}
			\begin{ruledtabular}
				\begin{tabular}{@{\,}l*{15}{c}@{\,}}
					TSLs & \thead{$ \mathbb{Z}_2 $ \footnote{$ \mathbb{Z}_2 : (\nu_0;\nu_1,\nu_2,\nu_3)$ } } & \thead{${C_m^{\rm {( \Gamma M K )}}}$} & \thead{${C_m^{\rm {(A L H )}}}$}&\thead{${C_m^{\rm{(A L \Gamma M)}}}$}\\
					\midrule	
					cat-cat ($ d\geq 8 $, $ l \in $ even) & (0;0,0,0)       & 4    &0&2		& & \\
					an-an ($ d\geq 8 $, $ l \in $ even) & (0;0,0,0)       & 2    &0&2		& &  \\
					cat-an ($ d\geq 10 $, $ l \in $ odd) & (0;0,0,1)       & 3    &1&2		& &  	\\			
				\end{tabular}
			\end{ruledtabular}
		\end{center}
		\label{T1}
	\end{table}

	\subsubsection{\label{subsec:berrycurv} \textbf{Berry curvature }}

	\begin{figure}[!htbp]
		\centering
		\includegraphics[width=0.49\textwidth]{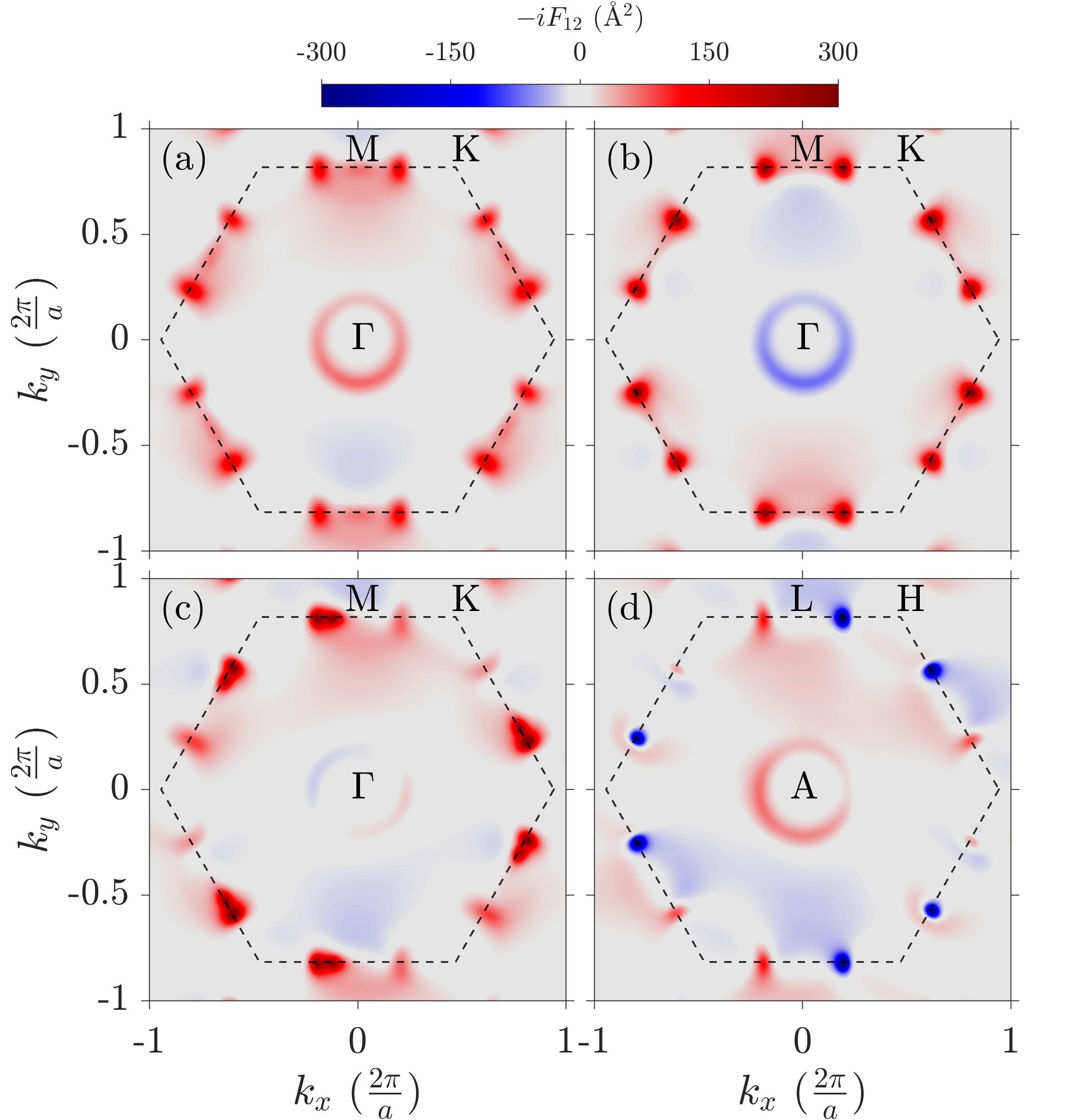}
		\subfloat{\label{fig:curv_cc}}				
		\subfloat{\label{fig:curv_aa}}
		\subfloat{\label{fig:curv_ca}}		
		\subfloat{\label{fig:curv_ca_up}}		
		\caption{\footnotesize  Berry curvatures associated with $ +i $ mirror subspace  calculated in the $ \rm \Gamma M K $ plane for (a) cat-cat and (b) an-an TSLs with 16 monolayer-height supercell, and for cat-an TSLs with 18 monolayer-height supercells in the (c)  $\rm \Gamma M K  $ and (d) $\rm  A L H  $ plane. Dashed lines denote the BZ boundaries.}\label{fig:curvature}
	\end{figure}
	
	The mirror Chern numbers $ C_m $ discussed in the last section were determined by integrating the Berry curvature for all occupied bands over the appropriate 2D cross sections of the 2D Brillouin zone, within $ +i $ mirror eigenstates.  We observe that the curvature field in the $ \rm \Gamma M K $ ($ \rm A L H $) plane is mostly concentrated in the vicinities of $\rm M $ ($\rm L $) and $\rm \Gamma $ ($\rm A $) points (see Fig.~\ref{fig:curvature}).  In particular, in Figure \ref{fig:curvature} we show the Berry curvatures for cat-cat/an-an and cat-an TSLs for $d=16$ and $d=18$, respectively. In cat-cat TSLs the Berry curvature extrema can be clearly discerned near $\rm M $ and $ \rm \Gamma $ points [Fig.~\subref*{fig:curv_cc}]. These extrema of curvature are analogously observed in an-an TSLs, but the sign of curvature turns out to be negative near the $ \rm \Gamma $ point  [Fig.~\subref*{fig:curv_aa}].  For alternating TSLs (cat-an), the Berry curvature features extrema of opposite signs near the $ \rm \Gamma  $ point, while uncompensated extrema near the $\rm M $ points are present [Fig.~\subref*{fig:curv_ca}]. In the $ \rm A L H $ plane positive Berry curvature is concentrated around $\rm A$, while near each $\rm L$ point there are two extrema of opposite signs.  In all cases, the high intensities of the Berry curvature in Fig.~\ref{fig:curvature} are correlated with the band edges shown in Fig.~\ref{fig:bs_t11_t22}.
	
	To quantify the above statements, we locally integrate the Berry curvature near high-symmetry points.   The calculated Berry flux is approximately $ 1 $  around each of the three $\rm  M $ points for all configurations, while at the $ \rm \Gamma $ point it is equal to $ 1 $, $ -1 $ and $ 0 $ for cat-cat, an-an, and cat-an TSLs, respectively. The sums of Berry fluxes near three $\rm M$ points and one $\Gamma$ point give the mirror Chern numbers obtained in Table~\ref{T1}.

	Interestingly, for the cat-an TSL in the $\rm ALH $ plane, the Berry flux is compensated near the $\rm  L $ points and amounts to $ 1 $ around the $\rm  A $ point. To explain how the results obtained for the $\rm ALH$ plane are different from those pertaining to $\rm \Gamma M K$ it is instructive to consider the mirror plane symmetry properties of the states on both planes. Let $\Psi_0$ and $\Psi_\pi$ be the Bloch functions defined on $ k_z = 0$ ($\rm \Gamma M K$ plane)   and $ k_z=\pi$ ($\rm ALH$ one). Let  $|\Psi_0 \rangle$ and $|\Psi_\pi \rangle$ be also  eigenstates of the (111) mirror plane operator.
	Then one can easily check that 
	\begin{equation} \label{eq:Mn0}
	\langle \Psi_0| \hat{M}_1 | \Psi_0 \rangle = \langle \Psi_0| \hat{M}_2 | \Psi_0 \rangle
	\end{equation}
	\begin{equation} \label{eq:MnPi}
	\langle \Psi_\pi| \hat{M}_1 | \Psi_\pi \rangle = - \langle \Psi_\pi| \hat{M}_2 | \Psi_\pi \rangle ,
	\end{equation}
	where $\hat{M}_1$ and $\hat{M}_2$ are mirror plane operators corresponding to $\rm TP_1$ and $\rm TP_2$. Let us assume that the influence on the topological properties of TSL depends mainly on individual TPs in the unit cell. Then we can define independent mirror-resolved Berry curvatures $F_{12}^{\mathrm{TP}_n,\pm i}$ and Chern numbers $C_{\mathrm{TP}_n,\pm i}$ determined individually for each ${\rm TP}_n$, where $\pm i$ denotes the eigenspace of the operator $\hat{M}_n$. The quantities defined in this way do not depend on the specific location of ${\rm TP}_n$ in the unit cell. We neglect the possibility that bulk states in the TSL also contribute to the total Berry curvature. The Chern numbers given in Table \ref{T1} and the curvatures shown in Fig. \ref{fig:curvature} are calculated with respect to the $\hat{M}_2$ operator. Thus, from equations \eqref{eq:Mn0} and \eqref{eq:MnPi} it follows that on the $ \rm \Gamma M K $ plane the total Berry curvature is a sum of curvatures $F_{12}^{\mathrm{TP_2},\pm i} + F_{12}^{\mathrm{TP_1},\pm i}$ determined by $\rm TP_2$ and $\rm TP_1$ (it is significant to remind that the TPs have different orientation, i.e., they are rotated by 180 degrees with respect to each other, so even if both are of the same kind, the two terms $F_{12}^{\mathrm{TP_n},\pm i}$ are different).  On the $ \rm A L H $ plane, the total Berry curvature is given by a different combination $ F_{12}^{\mathrm{TP_2},\pm i}\left(\vec{k}\right) + F_{12}^{\mathrm{TP_1},\mp i}\left(\vec{k}\right) = F_{12}^{\mathrm{TP_2},\pm i}\left(\vec{k}\right) - F_{12}^{\mathrm{TP_1},\pm i}\left(-\vec{k}\right)$, where in the last equality we used the fact that the $\pm i$ eigenstates of $\hat{M}_2$ at $\vec{k}$ and the $\mp i$ eigenstates at $-\vec{k}$ are related by time reversal. 
    In consequence, also the Chern numbers would be sums and differences of individual invariants: 
	\begin{equation} \label{eq:C_GMK}
	    C^\mathrm{(\Gamma MK)}_{\pm i} = C_{\mathrm{TP_1},\pm i} + C_{\mathrm{TP_2},\pm i},
	\end{equation}
	\begin{equation} \label{eq:C_ALH}
	C^\mathrm{(ALH)}_{\pm i} =  C_{\mathrm{TP_2},\pm i} -C_{\mathrm{TP_1},\pm i} ,
    \end{equation}
	where the subscripts in $C^\mathrm{(\Gamma MK)}_{\pm i}$ and $C^\mathrm{(ALH)}_{\pm i}$ denote the eigenspaces of the $\hat{M}_2$ operator. 
    
    This is consistent with the results compiled in Table~\ref{T1} if as individual Chern numbers we take $C_{\mathrm{cTP},m}=2$ for a cationic TP and $C_{\mathrm{aTP},m}=1$ for an anionic TP. As already mentioned, the curvatures shown in Fig.~\ref{fig:curvature} can be understood as the corresponding sums and differences of the curvatures determined by individual TPs. See also Fig.~\ref{fig:curvatureALH} in Appendix~\ref{app:BS} where the Berry curvature on the ALH plane for cat-cat and an-an TSLs are presented. The curvatures are much smaller than on $\rm \Gamma M K $ and integrate to zero, in line with the above observations.  
    
    We stress that Eqs.~\eqref{eq:C_GMK} and \eqref{eq:C_ALH} rely on quite strong assumptions on the wave functions in the TSL. Namely, we have postulated that the contributions to the band structure topology from the two TPs are well discerned, and also independent from each other and from all bulk electronic bands. Furthermore, we conjectured that the contribution of the latter is negligible. It is therefore necessary to confirm the appealing result of Eqs.~\eqref{eq:C_GMK} and \eqref{eq:C_ALH} in a separate calculation. For this purpose we will return to study independent topologies coming from individual TPs in section \ref{subsec:slab}.

	\subsubsection{\label{subsec:semi-infinite} \textbf{Surface states in  twinning superlattices}}

	\begin{figure*}[!htbp]
		\centering
		\includegraphics[width=0.9\textwidth]{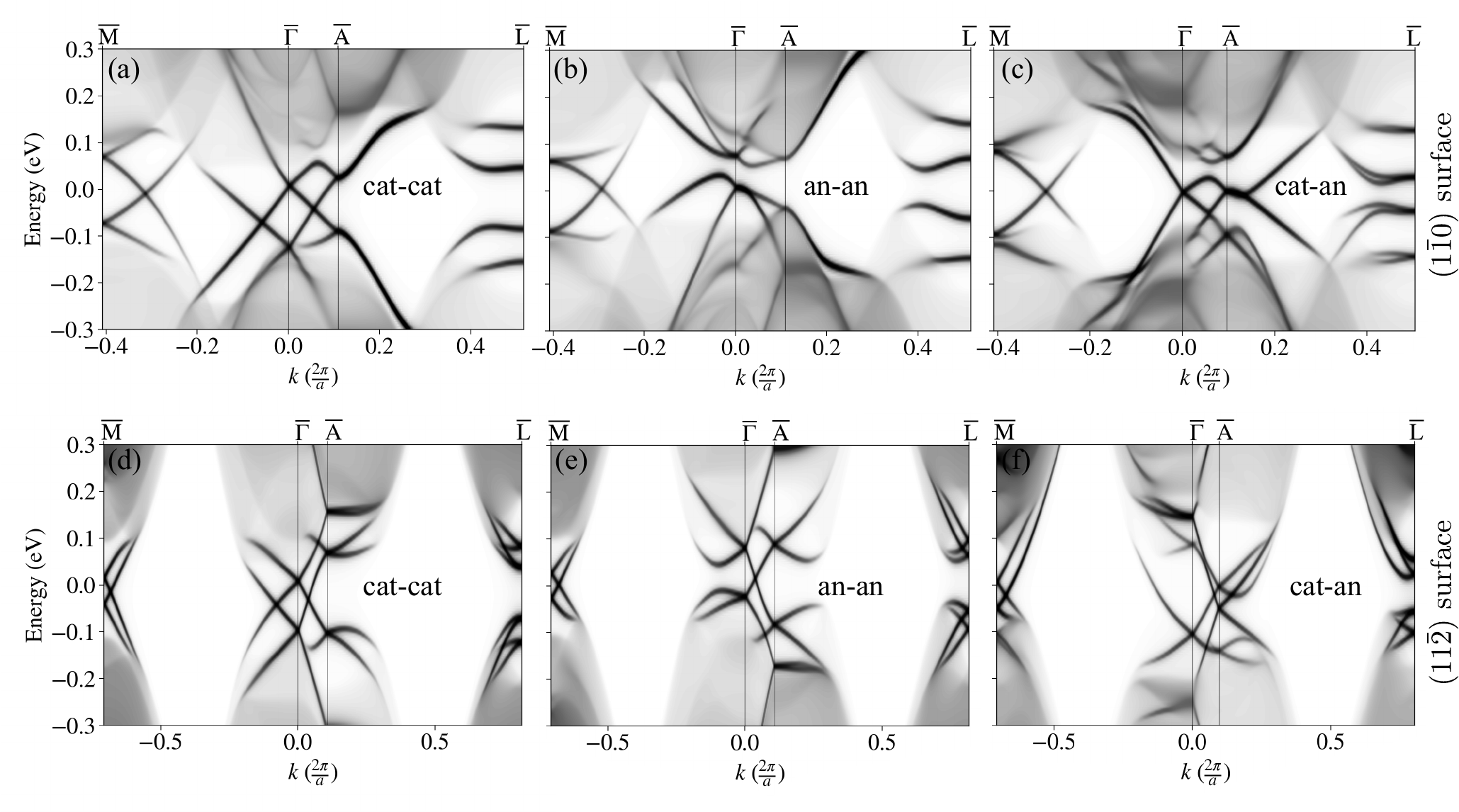}
		\subfloat{\label{fig:gf_cc}}				
		\subfloat{\label{fig:gf_aa}}
		\subfloat{\label{fig:gf_ca}}			
		\subfloat{\label{fig:gf_211_cc}}				
		\subfloat{\label{fig:gf_211_aa}}
		\subfloat{\label{fig:gf_211_ca}}		
		\caption{\footnotesize   Surface spectral functions of SnTe TSL  with $(1\bar{1} 0) $ (top row) and $(11\bar{2})$ (bottom row) surface orientations calculated with the iterative Green's function method. The spectra are obtained for $ d=16 $ supercell height for both cationic (a,d) and anionic (b,e) twin planes. (c) and (f) show the spectra for the $ d=18 $ supercell, which hosts one anionic and one cationic TP type.}\label{fig:spectral}
	\end{figure*}

	Searching for signatures of the topological invariants shown in Table~\ref{T1}, we investigate the lateral surface spectra of the TSLs. For calculations, we use the recursive Green's function method described in Ref.~\onlinecite{Sancho_1985} applied to the TB Hamiltonian of a semi-infinite TSL system terminated by a surface having one of two orientations, namely $ (1 \bar 1 0) $ and $ (11\bar{2}) $, which are perpendicular to the TSL growth axis. The projections of the high-symmetry points from the TSL BZ to the surface BZ for both orientations are shown in Fig.~\ref{fig:bz_hex}. The supercell height chosen for cat-cat and an-an TPs is $ d=16 $, while for TSLs with alternating TPs it is $ d=18 $. 
	The Berry curvatures in Fig.~\ref{fig:curvature} show that topological surface states can be expected to emerge in the close vicinity of the projections of $ \rm \Gamma $, $\rm  M $ and $\rm  A $ points from the 3D BZ, depending on the choice of TPs. Furthermore, $\rm  \overline{ M} \text{-}\overline{\rm \Gamma}$, $\rm  \overline{A}$-$\rm  \overline{L} $  and $\rm  \overline{ \rm \Gamma} $-$\rm  \overline{A } $ are the symmetry lines where the gapless surface states are predicted.
	
	Figures \subref*{fig:gf_cc}-\subref{fig:gf_ca} show the calculated spectral functions of a $ (1 \bar 1 0) $ plane for cat-cat, an-an, and cat-an TSLs, respectively. Plots for all types of TSL share the same feature of two separated-in-energy Dirac points at $\rm  \overline{ M} $ and two secondary Dirac points in the middle of the gap, which are shifted away from $\rm  \overline{ M} $ towards $\rm  \overline{\rm \Gamma}$ (only one is shown in the figure). The spectrum for cat-cat TSL [Fig.~\subref*{fig:gf_cc}] includes a similar structure near $ \overline{\rm \Gamma} $, with secondary Dirac points shifted towards $\rm  \overline{ M} $. In contrast, the spectrum of an-an TSL [Fig.~\subref*{fig:gf_aa}] is gapped in the vicinity of $ \overline{\rm \Gamma} $. In cat-an TSL [Fig.~\subref*{fig:gf_ca}], a single  topologically protected Dirac points appears exactly at $ \overline{\rm \Gamma} $ and at $ \rm \overline{ A}$.

	In Figs.~\subref*{fig:gf_211_cc}-\subref{fig:gf_211_ca} the calculated spectral functions of a $ (11\bar 2) $ surface are presented. The results along the $\rm  \overline{ M} \text{-}\overline{\rm \Gamma}$ line are qualitatively equivalent to the results for the $ (1\bar 1 0) $ surface. Moreover, along the $\rm \overline{ \Gamma} \text{-} \overline{A }$ line, a Dirac crossing of surface states is also observed. It results from the protection by the $ (1\bar 1 0) $ mirror symmetry which is not broken by the surface and the fact that $\rm  \overline{ \rm \Gamma} \text{-} \overline{A }$ line is a projection of the $ (1\bar 1 0) $ plane (see Fig.~\ref{fig:bz_hex}). This protection is the same as the protection provided by preserved $\left\lbrace 1 1 0 \right\rbrace $ symmetries in the non-defected TCI bulk. In the case of the $ (1\bar 1 0) $ TSL surface, all $\left\lbrace 1 1 0 \right\rbrace $ mirror symmetries are broken by the surface or by TPs and the states along $\rm  \overline{\rm \Gamma} \text{-} \overline{A} $ do not connect the valence band to the conduction band.
	
	The number of surface modes in the spectra is consistent with the calculated Chern numbers. The existence of topological surface states along the $(111)$ mirror symmetric lines is determined by the topologies of the electronic structures of the two kinds of TPs. Moreover, the presence or absence of Dirac crossings along $\rm  \overline{ \rm \Gamma}\text{-}\overline{ M} $, for both surface orientations, can be understood by inspecting the Berry curvatures in Fig.~\ref{fig:curvature}. The $\overline{ \rm \Gamma}$ point is in both cases a projection of both $\rm\Gamma$ and $\rm M$. For an-an TSLs, the Berry fluxes associated with these points have opposite signs, and their contributions cancel when projected on the surface, resulting in no topological states near $\overline{ \rm \Gamma}$. On the contrary, for cat-cat TSL, the contributions of the $\rm\Gamma$ and $\rm M$ valleys have matching signs, and each of the valleys generates a surface Dirac cone at $\rm\overline{\Gamma}$. In a cat-an TSL the vicinity of $\rm\Gamma$ is topologically trivial, and hence only the $\rm M$ valley produces a single Dirac cone on the surface. The $\rm \overline{ M} $ point for both surface orientations is a projection of two different $\rm M$ points, which by symmetry have matching Berry curvature profiles. Each of the $\rm M$ points in the TSL generates a Dirac cone at $\rm \overline{M}$ on the surface. We briefly note that the splitting in the energies of the Dirac points at $\rm \overline{ M}$ and, in cat-cat TSL, at $\rm\overline{ \Gamma}$ is due to valley mixing, i.e., interference between states coming from the two valleys projecting onto the same area in the surface BZ. 
	
    Similar observations can be made for the spectra along $ \rm \overline{ A}\text{-}\overline{ L}$. One Dirac point exists only in the cat-an case. It is situated at the $ \rm \overline{ A}$ point, which is consistent with nontrivial curvature around A and weak $\mathbb{Z}_2$ and mirror Chern numbers for the ALH plane.

\subsection{\label{subsec:slab}   (111) slab with a single twin plane}
	
	\subsubsection{\label{subsec:clean_slab}   \textbf{Slab with clean surfaces} }
	
	To conclusively determine the role of individual TPs, we turn to the problem of a twinned SnTe crystal with a single (111) twin boundary. As the lattice is locally mirror symmetric about the TP, we choose to investigate a finite system with a global (111) mirror symmetry. We consider the geometry of a (111)-oriented slab, with either a cationic or an anionic TP in the middle. The slab thickness of 121 atomic layers (ca. 21.8~nm) is chosen large enough to suppress any hybridization between states localized on the surfaces and on the TP.  Without loss of generality, we chose the slab surfaces to terminate with the same kinds of atoms as the ones that form the TPs.
	
	\begin{figure}[!htbp]
		\centering
		\includegraphics[width=0.48\textwidth]{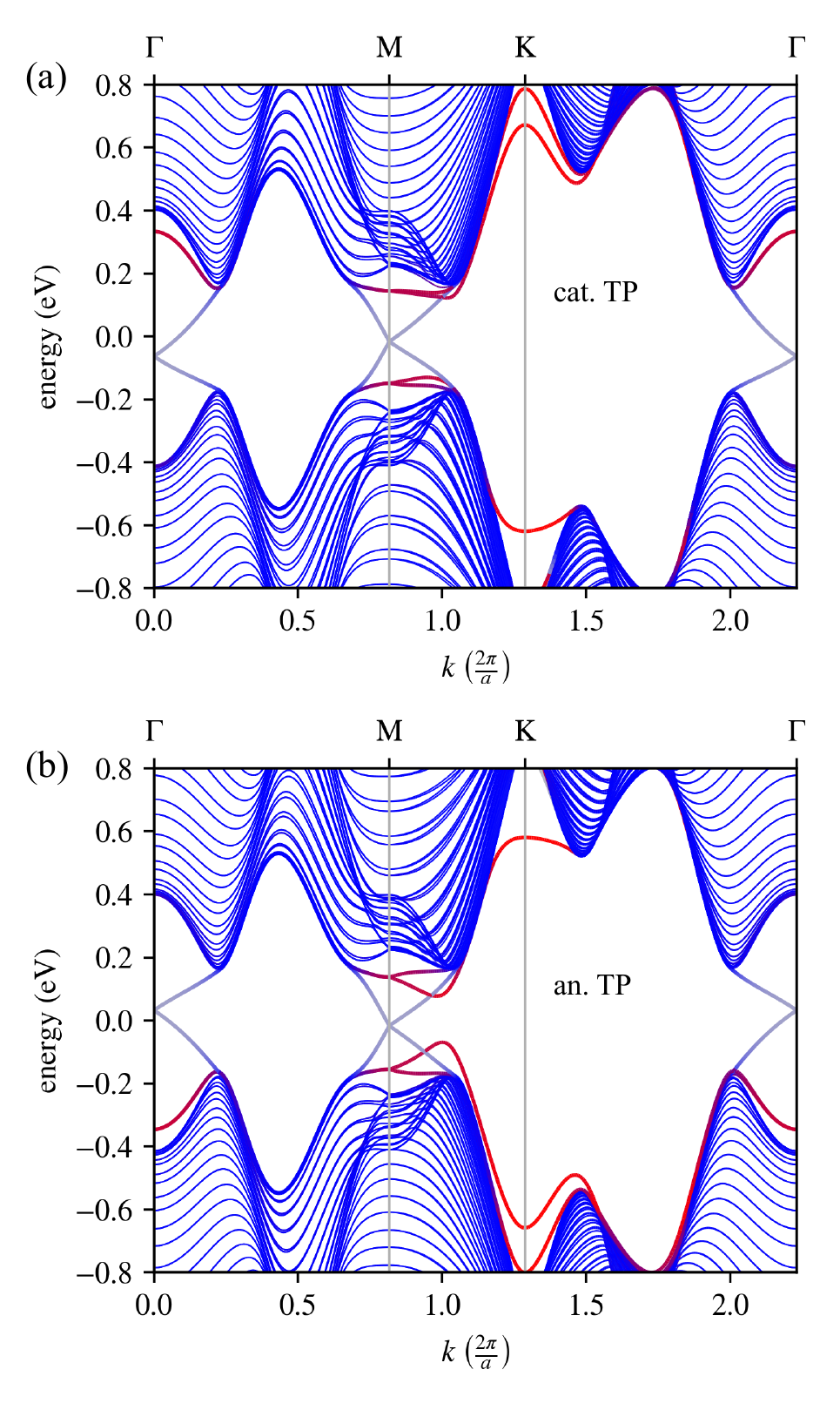}
		\caption{\footnotesize   The calculated band structures of [111]-oriented twinned SnTe slabs with 121 atomic layers (ca. 21.8~nm) thickness, featuring (a) a cationic or (b) an anionic TP in the middle, with (a) cationic and (b) anionic terminations.  The line color indicates that most of the weight of the wave function is located (red) near the TP, (grey) near the surfaces, or (blue) in the intermediate space.}\label{fig:bs_slabs}
	\end{figure}
	
	\begin{figure*}[!htbp]
		\centering
		\includegraphics[width=0.8\textwidth]{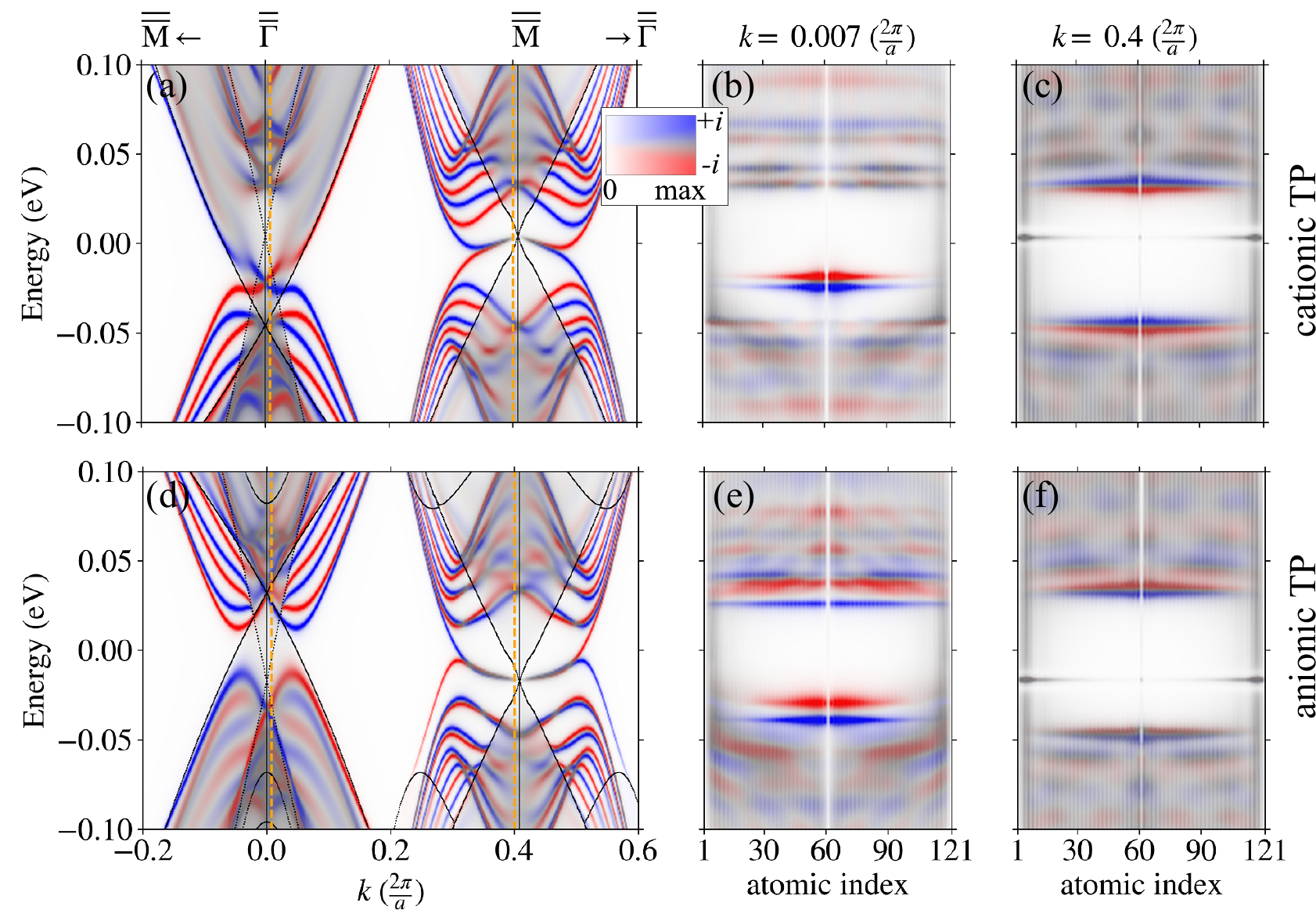}
		\subfloat{\label{fig:edge_c}}
		\subfloat{\label{fig:edge_c_G}}
		\subfloat{\label{fig:edge_c_M}}		
		\subfloat{\label{fig:edge_a}}
		\subfloat{\label{fig:edge_a_G}}		
		\subfloat{\label{fig:edge_a_M}}		
		\caption{\footnotesize   Edge states in 121-monolayer-thick (ca. 21.8~nm) (111)-SnTe slab along 1DBZ of its $ [11\bar2] $ edge. Calculated edge spectral functions for (a) a cation-terminated slab with cationic TP and (b) an anion-terminated slab with anionic TP. Panels (b)-(c) and (e)-(f) show the distributions of the spectral weight across the edge for different $k$ values, which are indicated by the vertical orange dashed lines in (a) and (d), for cationic and anionic TP, respectively. Outlines of the 2D slab bands projected to the edge are given by black dotted lines. The mirror-separated spectral functions of the subspaces $ +i $ and $ -i $ are encoded in blue and red, respectively. 
		}\label{fig:gf_edge}
	\end{figure*}	
	
	Figure~\ref{fig:bs_slabs} shows the spectra of the two kinds of slabs calculated along the $\rm \overline{\Gamma}\text{-}\overline{M}\text{-}\overline{K}\text{-}\overline{\Gamma}$ line.	For both cases, four topologically protected surface Dirac cones (one at the $\rm \overline{\Gamma}$ point and three at the $\rm \overline{M}$ points)  appear in the (111) slab BZ. For the purpose of better presentation, we added a positive (negative) onsite potential shift to the outermost layers of cationic (anionic) slab surfaces to shift the surface Dirac point to the middle of the spectrum. This change is irrelevant to the substance of our study.
	
	The spectral functions of the slab edge, calculated with the iterative Green's function method, are shown in Fig.~\ref{fig:gf_edge}.  The edge is oriented along $ [11\bar2] $ direction and it can be thought of as the $(1\bar{1}0)$ surface of the TSL, symmetrically truncated along $(111)$ atomic planes (see Figs.~\ref{fig:str} and \ref{fig:bz_hex}).  The spectral function map is color-coded to distinguish between the $ +i$ (blue) and $-i $ (red) mirror reflection eigenspaces. We show that in the case of a slab hosting a cationic TP, we observe two edge mode crossings [see Fig.~\subref*{fig:edge_c}]--one at $ \rm \overline{\overline{\Gamma}} $ with a Dirac dispersion and the second at $\rm  \overline{\overline{M}} $. The slab with an anionic TP has only one crossing located at the $\rm  \overline{\overline{M}} $ point, as shown in Fig.~\subref*{fig:edge_a}. In the cationic case, the spectral density of the edge states crossing near $\overline{\overline{\Gamma}}$ in Fig.~\subref*{fig:edge_c} is superimposed on the spectral density of the $\overline{\Gamma}$ and $\rm \overline{M}_2$ surface Dirac cones. However, we have verified that neither the dispersion nor the energy of these edge bands are affected when we change the surface states' energy with the added on-site potential on the outermost atomic layers. It can be seen that the wave functions that form the topological crossing at the $\rm \overline{\overline{\Gamma}}$ point are well confined in the [111] direction to the vicinity of the cationic TP [Fig.~\subref*{fig:edge_c_G}], while they are absent in case of anionic TP [Fig.~\subref*{fig:edge_a_G}].	Conversely, for both types of slabs, the edge modes near $\rm  \overline{\overline{M}} $ cross exactly at the (111) surface Dirac point corresponding to the $\rm \overline{M}_1$ and $\rm \overline{M}_3$ points in the slab BZ. This is accompanied by the change of the localization of edge state wave functions from tightly localized states at the TP to the localization close to the surfaces when $k$ approaches the $\rm  \overline{\overline{M}} $ point, as shown in Figs~\subref*{fig:edge_c},\subref{fig:edge_c_M} and \subref*{fig:edge_a},\subref{fig:edge_a_M}.

	\subsubsection{\label{subsec:local_BC}   \textbf{Band topology of individual TPs} }

	The Dirac cones at the slab surfaces can be gapped by adding a small $[111]$-directed Zeeman term $H_Z = m \sigma_z$ to the surface layers, where $m$ is the magnitude, and $\sigma_z$ is the third Pauli matrix acting in spin subspace. This breaks the time-reversal symmetry, but preserves the (111) mirror plane. Effectively, it can be interpreted as adding a weak magnetic field of equal magnitude and direction to the two surfaces. The Zeeman perturbation can introduce new topological properties of the surfaces and add new topological edge states. Nevertheless, we will show that this method allows the study of the Berry curvature and the calculation of the Chern number of the TP within the slab.

	To calculate the mirror Chern number $C_m$ of the slabs, the full Hamiltonian is first decomposed into $\pm i$ subspaces of the $(111)$ mirror reflection operator and the associated Berry curvatures $F_{12}^{\pm i}$ are calculated. Recall the mirror Berry curvature $F_{12}^m = (F_{12}^{+i}-F_{12}^{-i})/2$.	
	
	\begin{figure}[!htbp]
		\centering
		\includegraphics[width=0.49\textwidth]{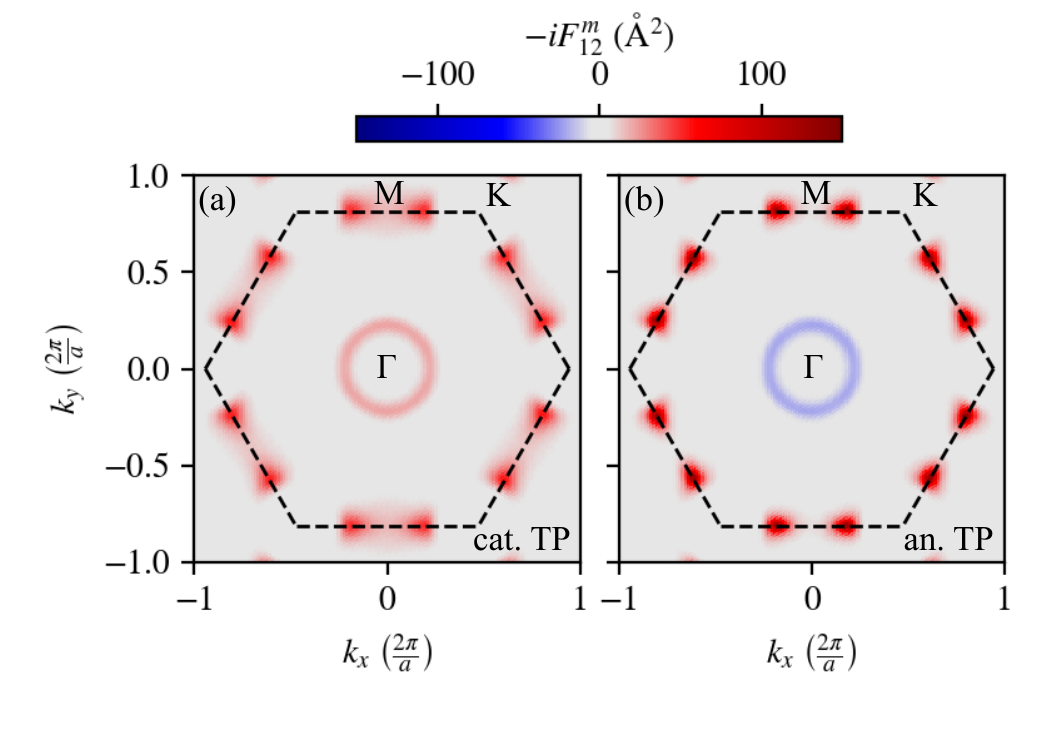}
		\subfloat{\label{fig:mirr_curv_c}}				
		\subfloat{\label{fig:mirr_curv_a}}
		\caption{\footnotesize    Mirror Berry curvatures of (111)-oriented twinned crystal slabs with (a) a cationic and (b) an anionic TP in the middle. The dashed lines denote the first BZ boundaries. A Zeeman term is added to the outermost layers to open the gap in the surface spectrum.	}\label{fig:mirror_Berry_slab}
	\end{figure}		
	
	Crucially, the surfaces do not contribute to $F_{12}^m$. By mirror symmetry, the wave functions on the top surface are related by a global (i.e., identical for all $\vec{k}$ points) unitary transformation to the wave functions on the bottom surface. Furthermore, in a sufficiently thick TCI slab, the wave functions at the two surfaces form two disjoint sets. Due to invariance of the Berry curvature to a $k$-independent unitary transformation, the Berry curvatures of the two surfaces are equal, $F_{12}^\textrm{top surf.}= F_{12}^\textrm{bottom surf.}$. It follows trivially that the surface contributions to the mirror-resolved Berry curvatures satisfy $F_{12}^{\mathrm{surf.},{+i}}=F_{12}^{\mathrm{surf.},{-i}}$, since the eigenfunctions of the (111) mirror symmetry operator pertaining to the surfaces have equal weights on both sides of the slab.	It is noted that the Zeeman terms on the surfaces affect the total Chern number $C$ through the total Berry curvature $F_{12}=F_{12}^{+i} + F_{12}^{-i}$, while the mirror Berry curvature $F^m_{12}$ contains information only on the TP and the surrounding twin lattice. In  Fig.~\ref{fig:mirror_Berry_slab}  the mirror Berry curvature maps of a cationic and an anionic twin boundary are shown. The mirror Berry fluxes around $\rm  \overline{M} $ and $ \rm \overline{\Gamma} $  points are fractional ($ \pm \frac{1}{2} $)  and eventually lead to mirror Chern numbers $C_m=2$ and $C_m=1$, for cationic and anionic TP, respectively.

	For the purpose of confining our analysis of the band structure topology to the vicinity of the TP, we introduce the projected Berry curvature
	\begin{equation} \label{eq:proj_Berry}
	F_{12}^{(A)}(\vec{k}) = \mathrm{Tr}[ \mathcal{F}(\vec{k}) P_A],
	\end{equation}
	where the trace is taken over all bands at $k$, $P_A$ is a projection operator corresponding to a selected part (labeled $A$) of the slab unit cell. A detailed description of the procedure of determining the projected Berry curvature is available in Appendix \ref{app:projection_F12}. Here we present an abridged explanation, excluding details not essential to our argument.

\begin{figure}[!htbp]
		\centering
		\includegraphics[width=0.49\textwidth]{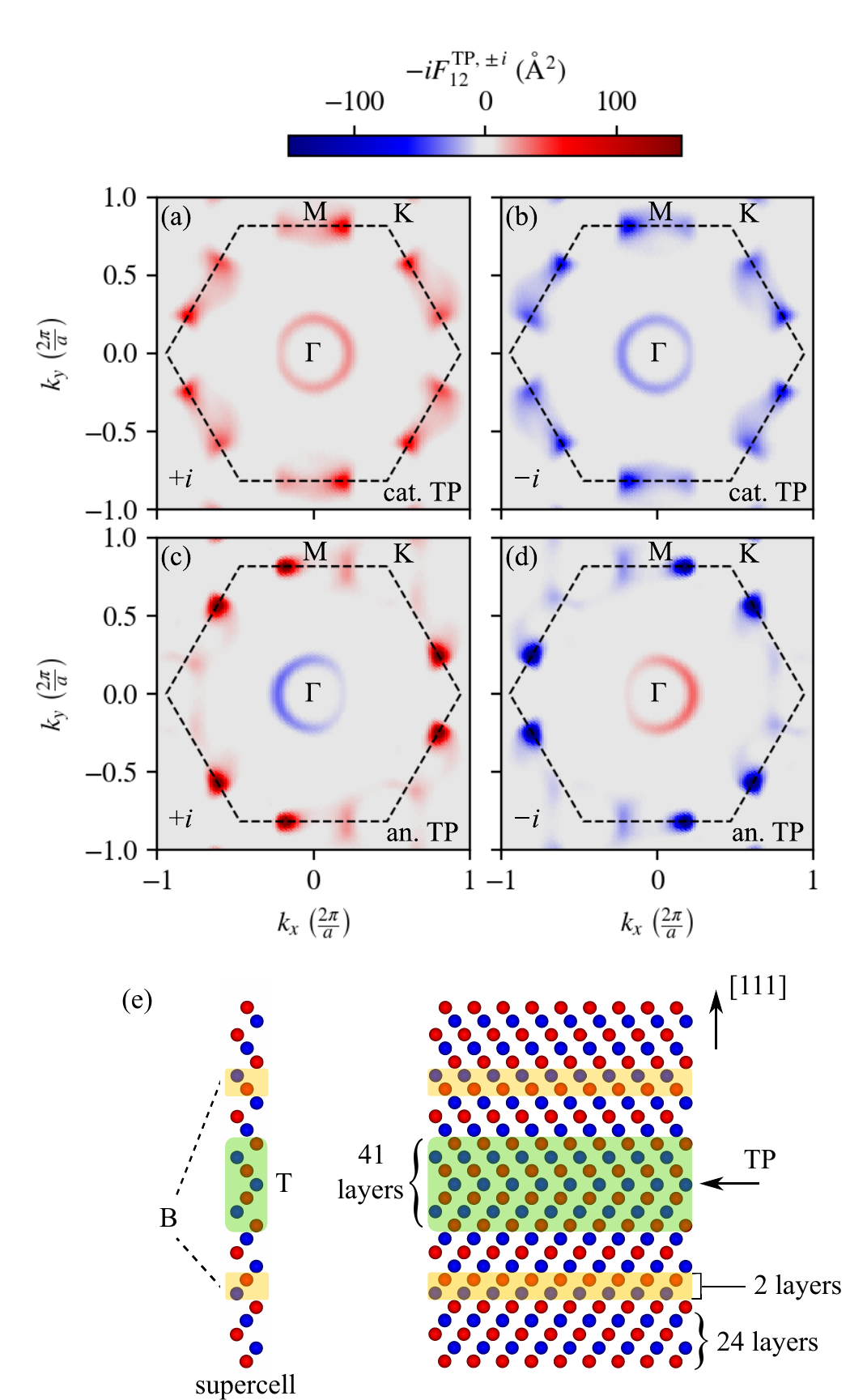}
		\subfloat{\label{fig:ITP_c_p}}				
		\subfloat{\label{fig:ITP_c_m}}				
		\subfloat{\label{fig:ITP_a_p}}				
		\subfloat{\label{fig:ITP_a_m}}				
		\subfloat{\label{fig:ITP_schematic}}				
		\caption{\footnotesize   The contribution to the Berry curvature of 121-monolayer thick, (111)-oriented twinned crystal slabs coming from the vicinity of  (a,b) the cationic and (c,d) the anionic TP, calculated in the (a,c) $+i$ and (b,d) $-i$ subspace of the (111) mirror symmetry operator. The dashed lines denote the first BZ boundaries.(e) Sketch of the slab with a schematic depiction of subsystems $T$ and $B$ that were used for the projected Berry curvature calculations.   }
		\label{fig:TP_curvatures}
	\end{figure}

    The Berry curvature of the slab can be split into three parts $F_{12}=F_{12}^\textrm{TP}+F_{12}^\textrm{surf.} + F_{12}^\textrm{bulk}$, where the subsequent addends arise from the presence of the TP, the surfaces, and the intermediate lattice, respectively. $F_{12}^\textrm{bulk}$ permeates the whole slab, including the surface and TP layers, while $F_{12}^\textrm{TP}$ and $F_{12}^\textrm{surf.}$ are confined to neighborhoods of their respective slab components. Accordingly, if $T$ denotes a range of atomic layers in the center of slab, containing the TP, $F_{12}^{(T)} = F_{12}^\textrm{TP} + (V_T/V) F_{12}^\textrm{bulk} $, where $V_{T}$ is the volume of $T$ and $V$ -- the volume of the entire slab. $F_{12}^\textrm{bulk}$ can be calculated by considering a range of layers $B$ of volume $V_B$ lying away from the TP and the surfaces, for which $F_{12}^{(B)} =  (V_B/V) F_{12}^\textrm{bulk} $. Thus, the contribution to the Berry curvature from the TP can be calculated as
    	\begin{equation} 
    	F_{12}^\textrm{TP} = F_{12}^{(T)} - \frac{V_T}{V_B} F_{12}^{(B)}.
    	\end{equation}
    A schematic depiction of volumes $T$ and $B$ is shown in Fig.~\ref{fig:ITP_schematic}.
	
	It's straightforward to show that the above scheme applies also to the projected curvatures $F_{12}^{\textrm{TP},\pm i},F_{12}^{(T),\pm i}, F_{12}^{(B),\pm i}$ calculated in the $\pm i$ subspaces of the mirror reflection operator.  We performed calculations of $F_{12}^{\textrm{TP},\pm i}$ for the two slabs described in Subsection~\ref{subsec:clean_slab}, with the magnetic term of magnitude $m=0.5$~eV added on the surfaces.  Through examining $F_{12}^{(A)}$ calculated for various parts $A$ of the slab, we establish that, in our tight-binding model, the appropriate choice of $T$ is 41 layers in the middle of the slab, while $B$ can be chosen as any two adjoining atomic layers (one cationic and one anionic), lying farther than 20 layers away from the TP and 24 layers away from the surface, along with two atomic layers being their mirror reflection on the other side of the TP. The maps of the calculated TP Berry curvatures are presented in Fig.~\ref{fig:TP_curvatures}, with panels (a,b) corresponding to the cationic TP, and panels (c,d) to the anionic TP. Panels (a,c) show the Berry curvatures in the $+i$ subspace, while panels (b,d) show the $-i$ subspace.
	
	For all calculated cases, the TP Berry curvature is concentrated around $\Gamma$ and $\rm M$. In analogy to the analysis of the TSLs in section~\ref{subsec:TSL} we calculate the TP Berry flux, by integrating $F_{12}^{\textrm{TP},\pm i}$ over vicinities of the high symmetry points. We find that for the cationic TP, it is ca. $\pm \frac{1}{2}$ in the $\pm i$ subspace, near both $\Gamma$ and $\rm M$. For the anonic TP, the TP Berry flux amounts to ca. $\mp \frac{1}{2}$ near $\Gamma$ and $\pm \frac{1}{2}$ near $\rm M$ in the $\pm i$ subspace. 
	
	Although the integral of $F_{12}^{\textrm{TP},\pm i}$ over the entire BZ is not quantized, the integer-valued Chern number of the $\pm i$ subspace can still be represented as $C_{\pm i} = c_{\textrm{TP},\pm i} + c_{\textrm{bulk},\pm i} + c_{\textrm{surf.},\pm i} $, i.e., the sum of real-valued components, obtained by integrating $F_{12}^{\textrm{TP},\pm i}$, $F_{12}^{\textrm{bulk},\pm i}$, and $F_{12}^{\textrm{surf},\pm i}$, respectively. However, by evaluating these integrals numerically, we find for the cationic TP $c_{\textrm{cTP},\pm i} \approx \pm 2$, and for the anionic TP $c_{\textrm{aTP},\pm i} \approx \pm 1$. In both cases $c_{\textrm{bulk},\pm i} \approx 0$. 
	This allows us to associate with the TPs an approximate mirror Chern number $c_{\textrm{TP},m} = (c_{\textrm{TP},+ i}-c_{\textrm{TP},- i})/2$, which is $c_{\textrm{cTP},m} \approx 2$ for the cationic TP and $c_{\textrm{aTP},m} \approx 1$ for the anionic TP.

	In conclusion, both types of twin boundaries act as 2D TCIs. Furthermore, as implied by the odd-valued  $c_{\textrm{aTP},m}$, a single anionic twin boundary is a 2D $\mathbb{Z}_2$ topological insulator, provided that no time-reversal symmetry breaking terms affect the vicinity of the TP.
	
	Our findings for individual TPs also have consequences for TSLs. The $+i$ Berry curvature of, e.g., a cat-an TSL on the $(\mathrm{\Gamma M K})$ plane, calculated with respect to the mirror plane lying in the anionic TP, is $F^{\textrm{TSL},(\mathrm{\Gamma M K}),+i}_{12} \approx F^{\textrm{aTP},+i}_{12} +F'^{\textrm{cTP},+i}_{12} + F^{\textrm{bulk},+i}_{12} $, where the prime symbol denotes rotating the map by $\pi$ around $[111]$. The rotation is due to the relative orientation of TP$_1$ and TP$_2$ in the TSL. On the $(\mathrm{ALH})$ plane, the analogous formula is $F^{\textrm{TSL},(\mathrm{ALH}),+i}_{12} \approx F^{\textrm{aTP},+i}_{12} +F'^{\textrm{cTP},-i}_{12} + F^{\textrm{bulk},+i}_{12} $. From time-reversal symmetry, it follows that $F'^{\textrm{cTP},-i}_{12}=-F^{\textrm{cTP},+i}_{12}$, which implies $F^{\textrm{TSL},(\mathrm{ALH}),+i}_{12} \approx F^{\textrm{aTP},+i}_{12} -F^{\textrm{cTP},+i}_{12} + F^{\textrm{bulk},+i}_{12} $. The application of the above reasoning to the $-i$ subspace and other TSL kinds is straightforward. This analysis explains very well the Berry curvature maps features in Figs.~\ref{fig:curvature} and~\ref{fig:curvatureALH} and shows that TPs in TSLs can be treated as almost independent systems.
	
	As a final note, we remark that although Berry flux around the three $\rm  \overline{M} $ points and the $ \rm \overline{\Gamma} $  point is fractional for a single TP, in Appendix~\ref{app:projection} we show that for any edge orientation these four points project onto the edge BZ in pairs, one from two $\rm  M $ points and the other from $ \rm \Gamma $ and $\rm  M $ points. Thus, Berry fluxes corresponding to the $\rm  M_i $ and $ \rm \Gamma $ points eventually sum to integers (1, $-1$ or 0) along the projection line.

	\section{\label{sec:conclusion} Conclusions}
	
	Our theoretical studies of the SnTe class TCI show that a TP, a 2D defect along [111] crystallographic direction, introduces a new two-dimensional topological system protected by the (111) mirror plane defined by the boundary. All presented in this paper calculations of topological invariants, number and position of the Dirac crossings through the energy gaps and the maps of Berry curvatures are consistent for TSLs and slabs. We can conclude that a TP defines a 2D TCI with mirror Chern number $C_m=1$ for an anionic TP, and $C_m=2$ for a cationic TP. We have also verified that similar calculations performed for TPs in a trivial insulator (PbTe) show that TPs are trivial. This means that the bulk topology due to the inverted gap at the $\rm L^{fcc}$ points is crucial for the nontrivial properties of TPs which arise from the introduction of a new mirror plane in an already topologically nontrivial bulk. 
		
	Calculations of the projected Chern number in thick slabs demand at least 41 atomic layers around the TP to converge. This means that the topology is defined not only by the components of the valence band wave functions well localized on the TP but also by the components delocalized in the bulk. On the other hand, it turned out that in TSLs the Chern numbers are well defined also for much smaller distances between TPs. This means that weak coupling between TPs does not destroy their topological properties and the topology of the TSL band structure can be determined from the sum of the properties of individual TPs.

	\begin{acknowledgments}
	    We acknowledge financial support by the Polish National Science Centre (NCN) Grant under project No. 2016/23/B/ST3/03725. Computations were carried out using the computers of Centre of Informatics Tricity Academic Supercomputer \& Network.
	\end{acknowledgments}
	\appendix

	\section{\label{app:BS}  Electronic structures along high-symmetry lines in $ \rm \mathbf{ALH} $ and $ \rm \mathbf{\Gamma ALM} $ planes}

	\begin{figure}[!htbp]
		\centering
		\includegraphics[width=0.4\textwidth]{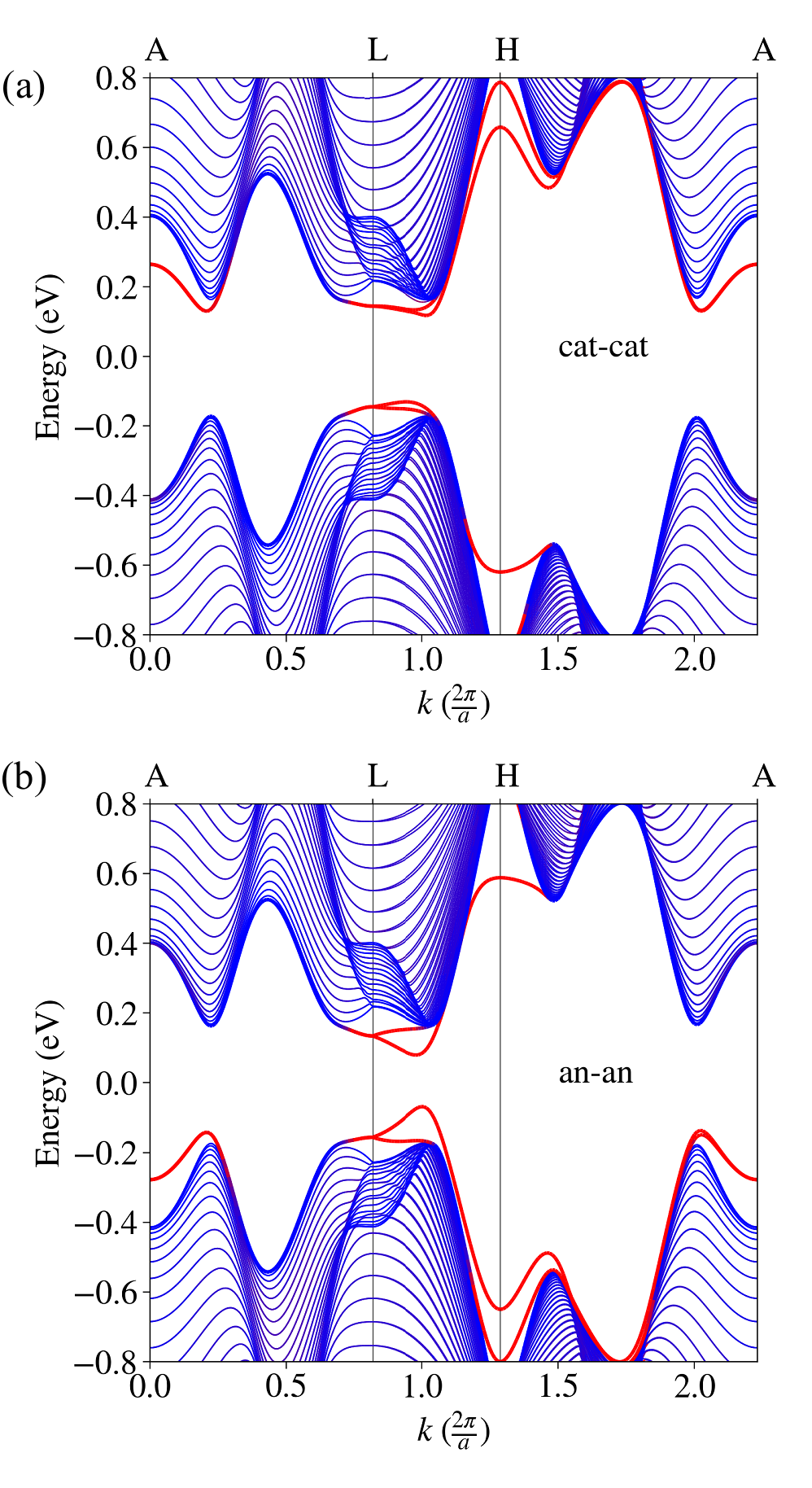}
		\subfloat{\label{fig:ALHcc}}				
		\subfloat{\label{fig:ALHaa}}	
		\caption{\footnotesize   The calculated band structures of a 200-monolayer-height [111]-oriented SnTe TSLs for the $ k $  wave vectors along the $\rm  A\text{-}L\text{-}H\text{-}A $ high-symmetry lines of the BZ. The TSLs constructed for (a) all-cationic TPs and (b) all-anionic TPs, and the red to blue color coding indicates the localization of the wave function on the TPs and bulk-like atoms.}\label{fig:ALH}
	\end{figure}		
	
	\begin{figure}[!htbp]
		\centering
		\includegraphics[width=0.49\textwidth]{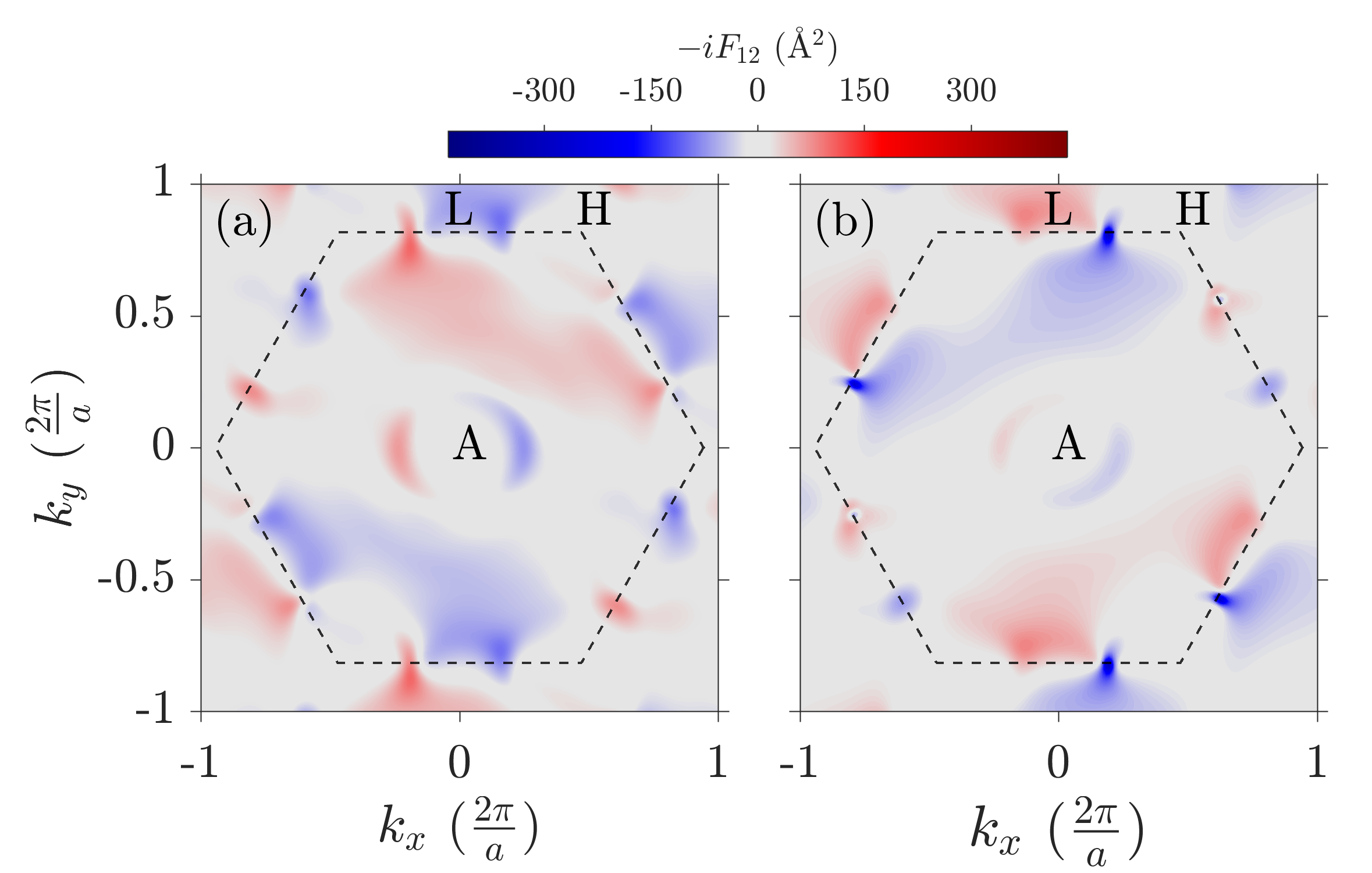}
		\subfloat{\label{fig:curv_ccALH}}				
		\subfloat{\label{fig:curv_aaALH}}
		\caption{\footnotesize Berry curvatures associated with the mirror subspace $ +i $ in the ALH plane of the 2D BZ. Figures~(a) and (b) correspond to cat-cat and an-an TSLs with 16 monolayer-height supercells, respectively. }
		\label{fig:curvatureALH}
	\end{figure}

	To support the results presented in the \ref{subsec:TSL} section, we further study the electronic structures of cat-cat/an-an TSLs along different paths in 3DBZ. The calculated spectra carried out for 200 monolayer supercell height are shown in Fig.~\ref{fig:ALH} for cat-cat/an-an TSL along the $\rm  A\text{-}L\text{-}H\text{-}A$ high-symmetry lines. The structures are very similar to those obtained in the $ \rm \Gamma MK $ plane for each type of TSL. In particular, the states localized near the TPs encoded by red in the figure have expectedly the same dispersion as in the $\rm  \Gamma MK $ plane. However, the calculated mirror Chern numbers presented in Table \ref{T1} are different between $\rm  \Gamma MK $ and $\rm  ALH $ and equal to zero in the latter plane for the TSLs that comprise a single TP type. This is due to the fact that upon mirror reflection the wave functions corresponding to the $\rm  ALH $ plane acquire a $ -1 $ phase factor difference between the two TPs in the supercell.  The consequences can be seen directly by calculating the Berry curvatures of the TSLs.  Therefore, we proceed with the Berry curvature calculations for the 16 monolayer height supercell for a cat-cat (an-an) TSL, as shown in Fig.~\subref*{fig:curv_ccALH} [Fig.~\subref*{fig:curv_aaALH}]. The curvature peaks for both types of TSL in the ALH plane are absent or are distributed oppositely around the $\rm  A $ and $\rm  L $ points. The Chern number amounts to zero within both mirror reflection subspaces $ +i $ and $ -i $.

\begin{figure}[!htbp]
		\centering
		\includegraphics[width=0.48\textwidth]{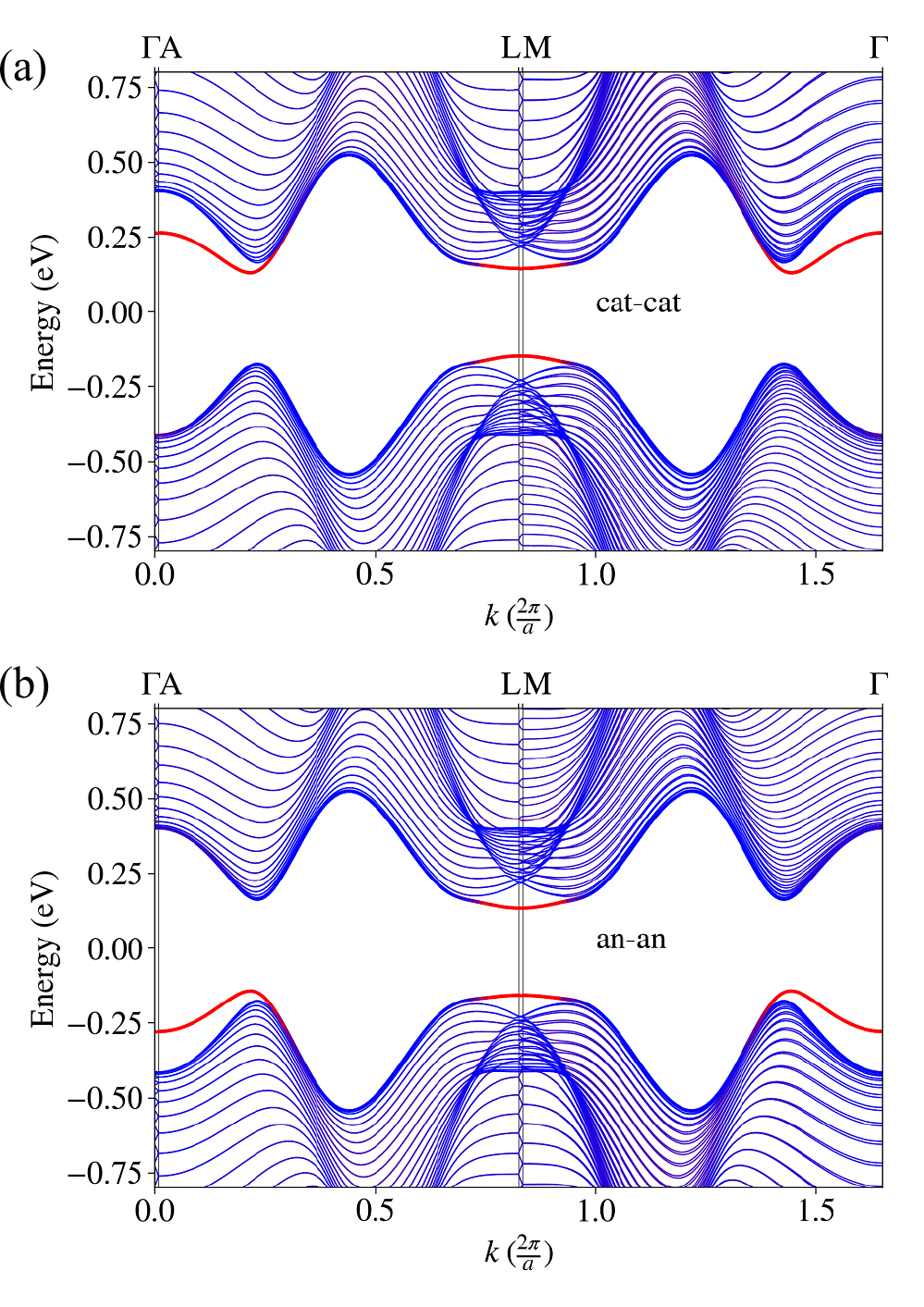}
		\subfloat{\label{fig:GALMGcc}}				
		\subfloat{\label{fig:GALMGaa}}						
		\caption{\footnotesize    The calculated band structures of a  [111]-oriented SnTe TSLs for the $ k $  wave vectors along $ \rm \Gamma $-$\rm  A $-$\rm  L $-$\rm M$-$\rm \Gamma $ high symmetry lines of the BZ. The 200 monolayer-height TSL supercells have (a) all-cationic and (b) all-anionic TPs. The red to blue color code indicates the location of the wave functions on TPs and bulk-like atoms.}\label{fig:GALMGall}
	\end{figure}

	Finally, we calculate the electronic structures along the high-symmetry lines that lie in the $ (1\bar 10) $ plane.  In Figs.~\subref*{fig:GALMGcc} and \subref*{fig:GALMGaa} we show the electronic structures of cat-cat and an-an TSLs, respectively. The supercells have a height equal to $ d=200 $ monolayers.  The length of the direction parallel to the [111] crystallographic axis is much smaller compared to the in-plane direction because the corresponding reciprocal lattice vector is inversely proportional to the height of the supercell.

	\section{\label{app:A_DiracPoint} Dirac cone at $\rm \overline{ A}$ point of the $(1\bar 10)$ surface of cat-an TSLs}
	
\begin{figure}[!htbp]
	\centering
	\includegraphics[width=0.4\textwidth]{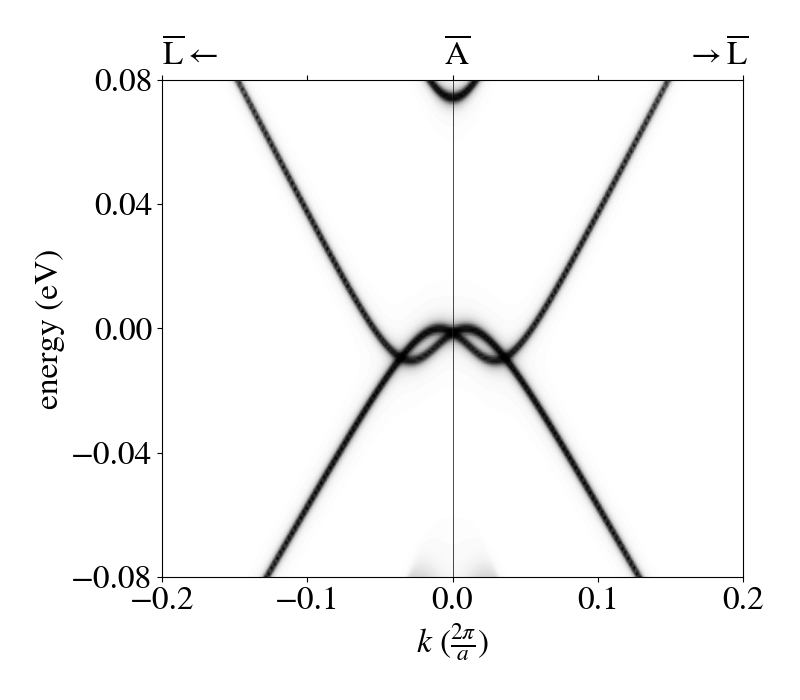}
	\caption{\footnotesize   Surface spectral functions of SnTe TSL  with $(1\bar{1} 0) $ surface orientation calculated with the $ d=18 $ supercell, which hosts one anionic and one cationic TPs.}\label{fig:ca_LAL}
	\end{figure}

	Cat-an TSLs feature nontrivial topology around the $\rm \overline{ A}$ point as is suggested by its Berry curvature in the ALH plane [see Fig.~\subref*{fig:curv_ca_up}]. Here, in Fig.~\ref{fig:ca_LAL} we show the spectral function of the $(1\bar 10)$ surface of the cat-an TSL along the $\rm \overline{ L}\textrm{-}\overline{ A}\textrm{-}\overline{ L}$ path in a closer view than the one demonstrated in Fig.~\subref*{fig:gf_ca}. The Dirac crossing at $\rm\overline{ A}$, consistently confirms the nontriviality of cat-an TSLs in the ALH plane with the mirror Chern number equal to $1$.

	\section{\label{app:projection_F12} Projected Berry curvature}
	In this section we describe more rigorously the definition of projected Berry curvature introduced in Subsection \ref{subsec:local_BC}. The projected Berry curvature was defined as
	\begin{equation} \label{eq:proj_Berry}
	F_{12}^{(A)}(\vec{k}) = \mathrm{Tr}[ \mathcal{F}(\vec{k}) P_A],
	\end{equation}
	where the trace is taken over all bands at $k$, $P_A$ is a projection operator corresponding to a selected part (labeled $A$) of the slab unit cell, and 
	\begin{align} \label{eq:Berry_operator}
	\begin{split}
	\mathcal{F}(\vec{k}) &=  \sum_{n,n',m} \frac{(-1)^{f_m}}{2} (f_m-f_n)(f_m-f_{n'})  \\
	&\ket{n}\left( \frac{\Braket{n|\frac{\partial H}{\partial k_x}|m} \Braket{m|\frac{\partial H}{\partial k_y}|n'}}{(E_m - E_n)(E_m - E_{n'})} \right. \\ 
	&\quad\left.- 
	\frac{\Braket{n|\frac{\partial H}{\partial k_y}|m} \Braket{m|\frac{\partial H}{\partial k_x}|n'}}{(E_m - E_n)(E_m - E_{n'})}
	\right)\bra{n'} \\
	\end{split}
	\end{align} 
	is an operator prepared such that $\mathrm{Tr}[ \mathcal{F}(\vec{k})]$ gives the Kubo formula for the Berry curvature. $H$ denotes the Hamiltonian of the slab, and $E_n$ the energy corresponding to the eigenvector $\ket{n}$ defining the $n$-th energy band at $\vec{k}$. $f_n$ is the occupancy, i.e. $f_n=1$ if the $n$-th band is occupied, and $f_n=0$ otherwise. $F_{12}^{(A)}$ is gauge-invariant and $\sum_A F_{12}^{(A)} = F_{12}$ as long as $\sum_A P_A = 1$. While~\eqref{eq:proj_Berry} is only one of many nonequivalent ways of decomposing Berry curvature into components corresponding to different subsets of the unit cell, we find it sufficient for our objective, as $F_{12}^{(A)}$ is sensitive to the weight on $A$ of the wave functions that contribute to the total Berry curvature.

	Since our objective is to classify the TPs by the mirror Chern number, the TP Berry curvatures have to be calculated separately for $\pm i$ subspaces of the $(111)$ mirror reflection operator. This is achieved by calculating
	\begin{equation}
	F^{ (A),\pm i}_{12}(\vec{k}) = \mathrm{Tr}[ P_{\pm i} \mathcal{F}(\vec{k}) P_{\pm i} P_A],
	\end{equation}
	which is effectively confining the sums in~\eqref{eq:Berry_operator} to eigenstates $P_{\pm i} \ket{n}$ belonging to the appropriate subspace. Note that $A$ must be chosen symmetric with respect to the $(111)$ mirror plane, ensuring $[P_{\pm i},P_A]=0$.

	\section{\label{app:projection} Projection rule of $\rm  M $ points in arbitrary edge of the 2D hexagonal lattice}

	Here, we demonstrate a general rule to determine whether the $\rm \Gamma$ point and $\rm  M $ points are always projected in pairs in any crystallographic direction. We note that an analogous argument also works for points $\rm A$ and $\rm L$.  Let us first define 2D primitive lattice vectors as $ \vec{t}_1 = a(0,\sqrt{3}) $ and $ \vec{t}_2=a(\frac{3}{2}, \frac{\sqrt{3}}{2}) $ to generate periodically infinite lattices in real space. The corresponding reciprocal lattice vectors are $ \vec{b}_1 = \frac{2\pi}{a} (-\frac{1}{3},\frac{\sqrt{3}}{3}) $ and $\vec{b}_2 = \frac{2\pi}{a} (\frac{2}{3},0)$, which define a 2D hexagonal BZ. The net vectors of the $\rm  M_i$ ($ i=1,2, 3 $) and $\rm \Gamma $ points are defined as follows.
	\begin{eqnarray}
	\Gamma = \vec{G},\\
	\vec{M}_1 = \frac{1}{2}(\vec{b}_1 + \vec{b}_2 ) + \vec{G},\\
	\vec{M}_2 =  \frac{1}{2}\vec{b}_2 + \vec{G},\\
	\vec{M}_3 =  -\frac{1}{2}\vec{b}_1 + \vec{G},
	\end{eqnarray}
	where $ \vec{G} $ is the reciprocal net vector $ n'\vec{b}_1 + m' \vec{b}_2 $ and $ n' $ and $ m' $ are integer numbers. The projection direction is along $ \vec{d}=-p \vec{t}_1 + q \vec{t}_2 $, which is perpendicular to the $ \vec{d}_\perp = p \vec{b}_2 + q \vec{b}_1$ direction.  We note that $ (p,q) $ being the edge indices are relatively prime. To obtain the projection in pairs, the following conditions must be satisfied.
	\begin{eqnarray}
	(\vec{M}_i - \vec{M}_j) \cdot \vec{d}_\perp =0, \qquad  \text{(pair of $ M $ points),}\\
	\vec{M}_k \cdot \vec{d}_\perp =0, \qquad  \text{(pair of $ \rm \Gamma  $ \& $\rm  M $ points).}
	\end{eqnarray}	
	After examination of all possible combinations of $ M $ points and using properties of the relatively prime numbers of the $ (p,q) $ edge, we conclude that the $\rm \Gamma$ point and the three $ M $ points are always projected in two separate pairs, namely:
	\begin{enumerate}
		\item $ (\rm M_2,M_3) $ and $ (\rm\Gamma, M_1) $, when both $ p $ and $ q $ are odd.
		\item $ (\rm M_1,M_2) $ and $ (\rm\Gamma, M_3) $, when $ p $ is even and $ q $ is odd.
		\item $ (\rm M_1,M_3) $ and $ (\rm\Gamma, M_2) $, when $ p $ is odd and $ q $ is even.
	\end{enumerate}

	\bibliography{ref}
	
\end{document}